\documentclass[a4paper, 12pt]{article}
\usepackage[dvips]{graphics}

\begin{document}

\title{\huge Evidence for a two scale Infrared NPQCD dynamics from  Pad\'{e} couplant structures}

\author{ F.  N.  Ndili\\
T.  W.  Bonner Nuclear Laboratory, MS315, Physics Department,\\
Rice University, Houston, TX.77251-1892, USA.}

\date{July, 2001}

\maketitle

\begin{abstract}
We report our finding of a two scale cut off structure in the infrared nonperturbative dynamics of Pad\'{e}
couplant QCD, in the flavor states $N_{f} \le 8$.  We argue that these two NPQCD momentum scales $Q_{c}(Y_{1})$
and $Q_{c}(Y_{2})$ can be identified as the scales of onset of chiral symmetry breaking and quark confinement
in QCD, and as the cut off boundaries of a strongly interacting infrared quark gluon phase of Quantum
Chromodynamics, intermediate between the hadronic phase of QCD and the weakly coupled perturbative QCD regime.  We  
investigsted the pattern of variation with flavor number $N_{f} \le 8$, of these domain boundaries of the intermediate
nonperturbative infrared QCD,  and found that the two scales are correlated, with a correlation factor
$Q_{c}(Y_{2})/Q_{c}(Y_{1})$ that rises to a peak at $N_{f} = 2$ from $N_{f} = 0$, but  falls off rapidly to zero for
$4 \le N_{f} \le 8$. We concluded firmly  that dynamical chiral symmetry breaking and quark confinement while being two
distinct QCD phenomena caused by two independent component QCD forces $Y_{1}$ and $Y_{2}$, are nevertheless
closely related phenomena of infrared nonperturbative QCD dynamics. Their correlation leads us to a finding that 
quark confinement is most favored in the $N_{f} = 2$ flavor  state of QCD, but becomes rapidly less probable
for $N_{f} \geq 4$, this finding being  exactly as one observes  in nature.

\end{abstract}

{\it{Keyword: Scales of Chiral symmetry breaking and confinement in QCD}\/}\\
{\bf PACS: 12.38.-t} \\
 E-mail: ndili@hbar.rice.edu

\newpage

\section{INTRODUCTION}
Quantum Chromodynamics (QCD) is known to have three main regimes or phases: (1) the perturbative QCD (PQCD) regime at
high momentum transfers $Q$ where the color force dynamics is weak and tends to zero for $Q \rightarrow \infty$; (2) the 
hadron regime at very low momentum transfers (the far infrared) where the color force has already grown so strong that the
quarks and gluons are confined inside hadrons and cease to be the observable degrees of freedom of QCD dynamics. (3) Then 
there is the intermediate (infrared) momentum transfer regime we shall here denote by CQCD, where the quarks and gluons
are still  the primary physical degrees of freedom of the QCD dynamics, but have their coupling constant 
sufficiently large that all the QCD processes in this regime are nonperturbative. These nonperturbative
processes include in particular chiral symmetry breaking and the  confinement of  quarks and gluons
into hadrons.  As conceived, the CQCD regime of QCD necessarily interfaces with the regime of applicability
of Chiral Perturbative theories and Effective Lagrangians~\cite{Leutwyler94, Weinberg79, Nambu61} of QCD, 
where Nambu-Goldstone bosons (pions, kaons..) formed from nonperturbative chiral symmetry breaking,  are  explicitly 
highlighted as the effective degrees of freedom in terms of which to formulate  effective Lagrangians of QCD dynamics
in this low energy region.

 The exact momentum scale points and critical couplings at which the CQCD regime sets in from the PQCD end, and  
later cuts off at the hadron end, are not known or measured, although  a number of phenomenological studies
particularly of chiral symmetry breaking, quark antiquark condensates, and quark 
confinement~\cite{Miransky93,Higashijima91, Manohar84, Shuryak81},
have placed some bench marks on the onset and cessation of  this CQCD regime of Quantum Chromodynamics.
Additionally, recent heavy ion experiments at CERN and Brookhaven (RHIC), approaching the same CQCD regime of QCD
from the hadron end (at high temperatures), have  confirmed the qualitative existence of this third phase of QCD 
called there Quark Gluon Plasma (QGP) phase, and  begun to  yield some quantitative information on the 
points of onset of deconfinement and chiral symmetry restoration in high temperature QCD. Finally, Lattice 
Quantum field theory simulations of QCD~\cite{Baal98, Bhattacharya2001} have  several hints of chiral symmetry
breaking and quark confinement phases in QCD, and have placed their own bench marks on the domain of the QGP
intermediate phase of QCD.

These known features of QCD, in particular the  three phase structure of QCD, lead us to consider seriously  
the recent finding in our paper~\cite{ndili2001}, based on  Pad\'{e} improved PQCD,  that the same low energy
region of QCD is governed by four component coupling constant solutions $a_{1}, a_{2}, a_{3}, a_{4}$,
(with $a_{i} = \alpha_{s}^{i}/ \pi),$ in which
$a_{1}$ and $a_{2}$ account for any pole structure or else coupling constant freezing in the low energy region of
QCD, while $a_{3}$ and $a_{4}$ provide a two scale cut off structure of PQCD from the NPQCD infrared region. The
possibility then arises that these two IR cut off scales $a_{3}$ and $a_{4}$ observed in our Pad\'{e} couplant QCD,
may be identified  as the onset points or boundaries of the CQCD regime of QCD, and in particular as the critical 
points of chiral symmetry breaking and quark confinement in QCD. This possibility is what we examine in this
paper, matching our observed two scale  Pad\'{e} infrared NPQCD coupling constant values and features $(a_{3}, a_{4})$,
with various existing  phenomenological bench marks and features of low energy QCD dynamics.  Because no such two scale
infrared cut off structure has previously been observed or reported in perturbative QCD studies with renormalization
group equations, in the infrared region, our work appears to us an interesting  new finding.

We present our study and results as follows.  In section 2, we recall some essential equations and features  of our
optimized $[1|1]$ Pad\'{e} QCD from our earlier paper~\cite{ndili2001}, noting in particular the observed
two scale $Q_{c}(Y_{1})$ and $Q_{c}(Y_{2})$ infrared cut off structure in the flavor states $N_{f} \le 8$ of our
optimized Pad\'{e} couplant QCD.  Because this earlier observation was based on the one physical observable 
or effective charge $R_{e^+e^-}(Q)$ and its renormalization scheme invariant $\rho_{2}$, 
we devote sections 3 and 4  of the paper, to considerations of what  these Pad\'{e}  couplant 
features look like, when other QCD physical observables, both timelike and spacelike, are used besides
$R_{e+e-}(Q)$. Satisfied that  the two scale infrared cut off feature persists in all QCD physical observables, 
timelike and spacelike,  we devote section 5 of the paper, to comparing this two scale feature with the known
phenomenological results  and bench marks of chiral symmetry breaking and confinement in QCD.  

Prompted by the good agreement we found, we devote section 6  of the paper to a consideration of the flavor dependence
of our infrared two scale structure and the implication of this for quark confinement in QCD.  Section 7 considers similar
implication for the question of infrared fixed point in low flavor states of QCD. The paper concludes with a brief summary 
in section 8.

\section{FEATURES OF THE OPTIMIZED $[1|1]$ PAD\'{E} QCD WITH  $R_{e+e-}(Q)$ OBSERVABLE.}
In our earlier paper~\cite{ndili2001}, we showed that the optimized $[1|1]$ Pad\'{e} improved PQCD couplant equation
bifurcates into two independent $Y_{1}$ and $Y_{2}$ couplant equations given by 

\begin{equation}\label{eq: ndili1}
\rho_{1}(Q, N_{f})  -  P_{1}  =  \frac{1}{\bar a}  + c \ln \left | \frac{ c \bar a}{1 + c \bar a  -  2 \bar a P_{1}} \right  |  
\end{equation}

\begin{equation}\label{eq: ndili2}
\rho_{1}(Q, N_{f})  -  P_{2}  =  \frac{1}{\bar a}  + c \ln \left | \frac{ c \bar a}{1 + c \bar a  -  2 \bar a P_{2}} \right  |  
\end{equation}

where:
\begin{equation}\label{eq: ndili3}
\rho_{1}(Q, N_{f}) = b \ln \frac{Q}{\Lambda_{\mathrm{QCD}}} + c \ln \frac{2c}{b} - r_{1} = \tau - r_{1}
\end{equation}

\begin{equation}\label{eq: ndili4}
P_{1}  = \frac{3c}{2}  + \frac{1}{2} \sqrt D  =  \frac{1}{2c} \bar c_{2}(+)               
\end{equation}

\begin{equation}\label{eq: ndili5}
P_{2}  = \frac{3c}{2}  - \frac{1}{2} \sqrt D  =  \frac{1}{2c} \bar c_{2}(-)                    
\end{equation}

and

\begin{equation}\label{eq: ndili6}
 D = 8c^2 - 4 \rho_{2}.
\end{equation}

Here $\bar a = \bar \alpha_{s}/ \pi$, is the optimized couplant solution of eqn.~(\ref{eq: ndili1})
or~(\ref{eq: ndili2}), while $\bar \alpha_{s}$ is the corresponding QCD coupling constant.  We shall from here
onwards drop the bar over both quantities purely as a matter of notational convenience but it is understood that 
all our computations and plots below refer to $\bar a$ or $\bar \alpha_{s} / \pi$.  The quantities
$b, c, c_{2}(\overline{MS})..$ are perturbative  QCD beta function coefficients  having  their 
usual values~\cite{Gross73,Caswell74,Tarasov80,Ritbergen97}:
\begin{equation}\label{eq: ndili7} 
b = \frac{33 - 2N_{f}}{6}   = 2 \beta_{0}                                         
\end{equation}
\begin{equation}\label{eq: ndili8} 
c = \frac{153 - 19 N_{f}}{2(33 - 2 N_{f})}   = \beta_{1}/ \beta_{0}                          
\end{equation}
\begin{equation}\label{eq: ndili9}
c_{2}(\overline{MS}) = \frac{3}{16(33 - 2 N_{f})} \left [ \frac{2857}{2} - \frac{5033}{18} N_{f} + \frac{325}{54} N_{f}^2  \right ] 
\end{equation}

$c_{3}(\overline{MS}) = \beta_{3}/ \beta_{0}$, where
\begin{equation}\label{eq: ndili10} 
\beta_{3} = 114.23033 - 27.133944 N_{f}  + 1.5823791 N_{f}^2 + 5.85669582 \times 10^{-3} N_{f}^3                             
\end{equation}

In the optimization process~\cite{ndili2001}, the coefficient $c_{2}$ as a variable satisfies a quadratic 
equation given by:

\begin{equation}\label{eq: ndili54}
c_{2}^2  -  (16c^2) c_{2}  + (4 c^2 \rho_{2}  + c^4)   =  0                          
\end{equation}

whose solutions are:

\begin{equation}\label{eq: ndili55}
c_{2} \rightarrow \bar c_{2} = 3c^2 \pm  c \sqrt D  =  2 c \left ( \frac{3}{2} c  \pm  \frac{1}{2} \sqrt D \right )  
\end{equation}

It follows from this that if the quantity $D$ is negative, $\bar c_{2}$ is not real and our $Y_{1}$ and $Y_{2}$ 
equations have no real solutions unless we work with $|D|$.   Since  $D =  8c^2 - 4 \rho_{2}$, this viability of
of eqn.~(\ref{eq: ndili54}) depends in turn on  the sign and magnitude of the renormalization scheme (RS)
invariant $\rho_{2}$.

These RS invariants $\rho_{1}, \rho_{2}, \rho_{3}....$ arise in general~\cite{Stevenson81,Stevenson86,Dhar83,Dhar84} from 
a consistency requirement between the PMS/ECH constraints on a perturbatively truncated generic physical observable

\begin{equation}\label{eq: ndili11}
R^{(n)}(Q) = a(1 + r_{1} a + r_{2} a^2 + ... )                                              
\end{equation}
and the basic renormalization group equation (RGE) for the couplant $a$.  Explicitly, the RS invariants are given by: 
 
\begin{equation}\label{eq: ndili12}
\rho_{1} (Q, N_{f})   =  \tau  -  r_{1}                                            
\end{equation}

\begin{equation}\label{eq: ndili13}
\rho_{2} (N_{f})  =  r_{2}  +  c_{2}  -  (r_{1}  + \frac{1}{2} c)^2                         
\end{equation}
or simply
\begin{equation}\label{eq: ndili13a}
\rho_{2} (N_{f})  =  r_{2}  +  c_{2}  -  r_{1}^2  -c r_{1}                         
\end{equation}

Their explicit numerical values and sign  then depend on the particular physical observable and its loop order coefficients
$(r_{1}, r_{2}, r_{3}. . ..)$ we use in evaluating the RS invariants.  In the particular case of the $R_{e+e-}(Q)$
observable used in paper~\cite{ndili2001}, the $\rho_{2}$ values are large and negative and eqns~(\ref{eq: ndili1})
and~(\ref{eq: ndili2}) have four crossing point solutions: $Y_{1} = a_{1}, a_{2}, a_{3}$ and $Y_{2} = a_{4}$. 

\begin{figure}
\scalebox{1.0}{\includegraphics{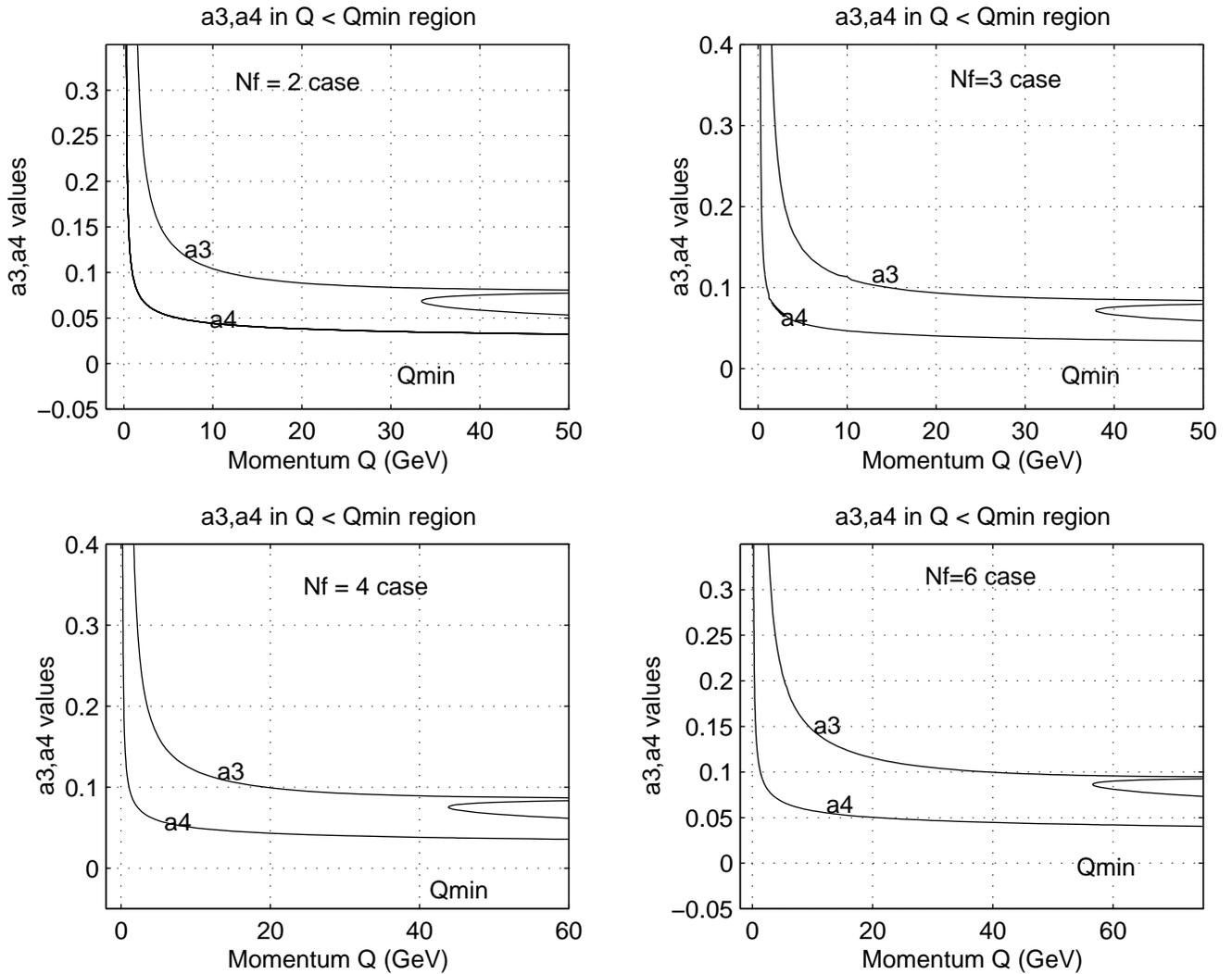}}
\caption{A plot from the $R_{e+e-}(Q)$ case, showing  that Pad\'{e} couplant solutions $a_{3}$ and $a_{4}$ in general
remain perturbative in the region $Q < Q_{\mathrm{min}}$ and for all flavors $N_{f} \le 8$, except close to their Landau 
type cut off point.}
\label{fig: ndili600}
\centering
\end{figure}

\begin{figure}
\scalebox{1.0}{\includegraphics{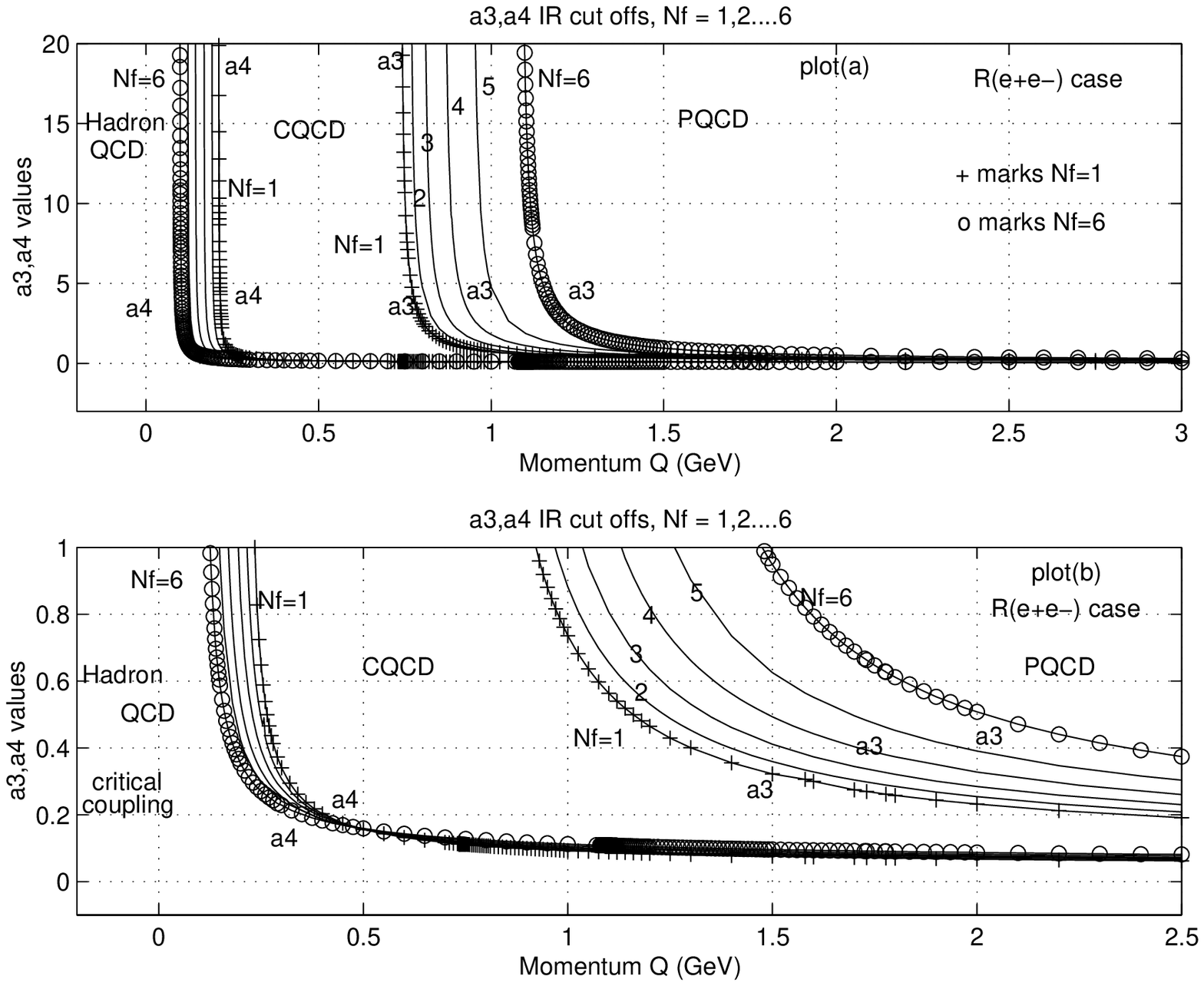}}
\caption{The infrared cut off points of Pad\'{e} couplant solutions $a_{3}$ and $a_{4}$ for various
flavors $N_{f} \le 8$, computed with respect to the $R_{e+e-}(Q)$ observable (effective charge)}
\label{fig: ndili600a}
\centering
\end{figure}

\begin{figure}
\scalebox{1.0}{\includegraphics{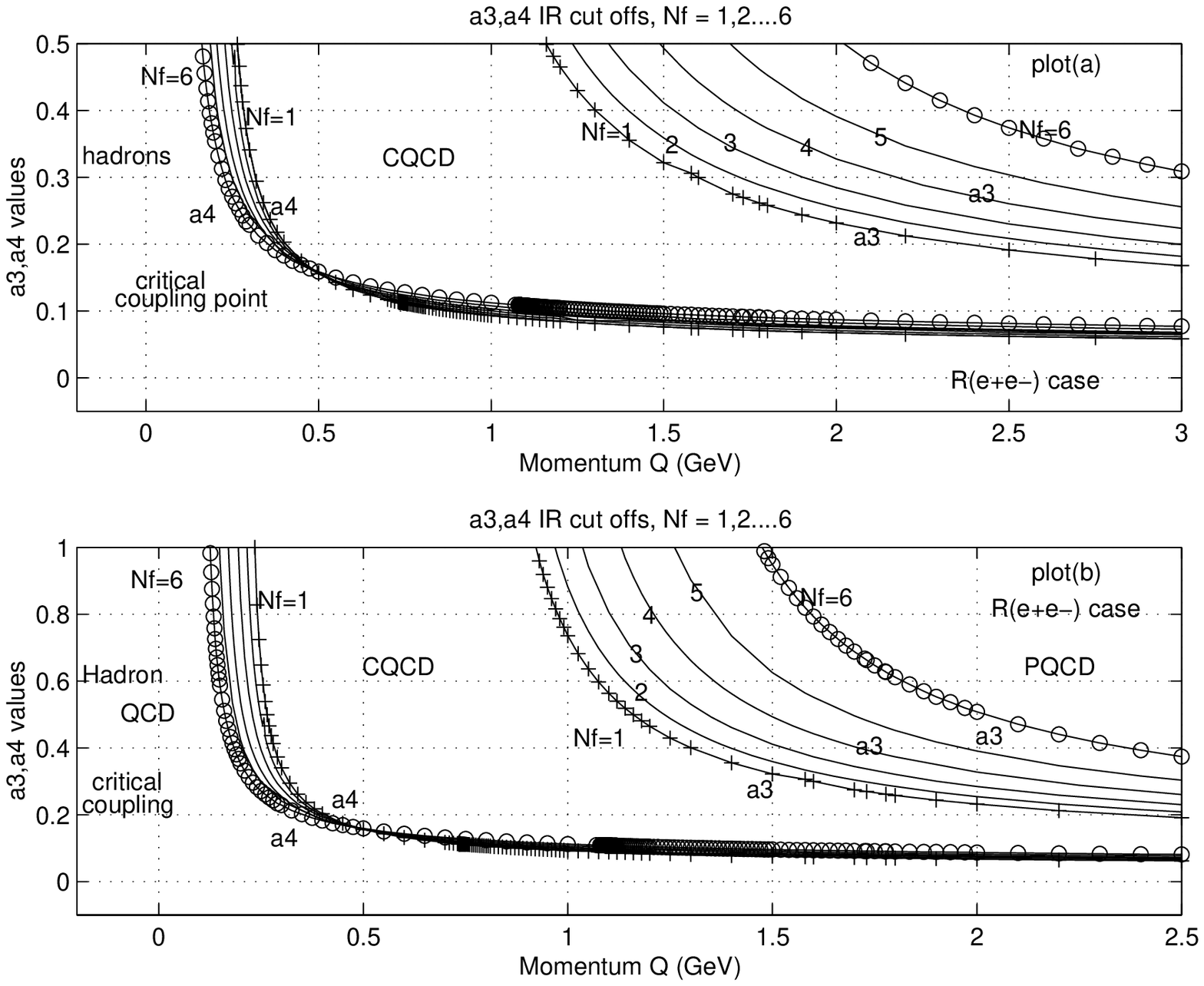}}
\caption{ Further view of the infrared cut off points of Pad\'{e} couplant solutions $a_{3}$ and
$a_{4}$ for various flavors $N_{f} \le 8$, computed with respect to the $R_{e+e-}(Q)$ observable (effective charge)}
\label{fig: ndili600b}
\centering
\end{figure}

Our finding in~\cite{ndili2001} is that solutions $a_{1}$ and $a_{2}$ determine an infrared attractor pole  behavior of
Pad\'{e} QCD for $N_{f} \le 8$, while  the same $a_{1}, a_{2}$ solutions determine an infrared fixed point  frozen 
couplant behavior for $N_{f} \geq 9$. On the other hand, the $a_{3}$ and $a_{4}$ solutions determine the points of cut
off of PQCD from the infrared region.  In the particular case of $N_{f} \le 8$ where $a_{3}$ and $a_{4}$ remain 
perturbative long after solutions $a_{1}$ and $a_{2}$ have turned away from the infrared region,  shown typically in 
fig.~\ref{fig: ndili600}, the ultimate cut off points $Q_{c}(Y_{1})$ of $a_{3}$ and $Q_{c}(Y_{2})$ of $a_{4}$ can become 
phyically meaningful, being two possible scales of NPQCD infrared dynamics that our Pad\'{e} couplant QCD predicts.
In this respect, we found earlier in paper~\cite{ndili2001} that while $a_{3}$ cuts off typically at a critical 
momentum $Q_{c}(Y_{1}) \approx 1.50$ GeV, the $a_{4}$ cuts off typically at
$Q_{c}(Y_{2})  \approx \Lambda_{\mathrm{QCD}} \approx 300$ MeV, shown in figs.~\ref{fig: ndili600a} 
and~\ref{fig: ndili600b}. The suggestion is that these two cut off points (and their corresponding critical
coupling constants) may be marking the domain of CQCD, such that we can
write: $Q_{c}(Y_{2})  \le  Q_{\mathrm{CQCD}} \le Q_{c}(Y_{1})$, with $Q_{c}(Y_{1}) = \Lambda_{\chi}$ being 
point of onset of chiral symmetry breaking, while $Q_{c}(Y_{2}) = \Lambda_{\mathrm{con}}$ is the critical scale point of
confinement and phase transition into hadrons.

Before however drawing this conclusion, we need to examine two points.  The first is to determine that the above
$a_{3}, a_{4}$ optimized Pad\'{e} two scale infrared couplant feature persists whatever the physical observable
and its renormalization scheme invariants $\rho_{i}$ we use, particularly spacelike and timelike physical observables.
The second point is to determine that  our numerical values of $Q_{c}(Y_{1})$ and $Q_{c}(Y_{2})$ from various
physical observables and $\rho_{2}$ values, timelike and spacelike, agree with and are consistent with what one has come to expect
from various existing  phenomenological studies of chiral symmetry breaking and confinement in QCD.  We now consider 
these two issues.

\section{VARIOUS SPACELIKE AND TIMELIKE QCD OBSERVABLES AND THEIR  INVARIANTS $\rho_{1}, \rho_{2}.$}
As already indicated above, while optimization and the entire effective charge formalism~\cite{Stevenson81,
Grunberg84,Kazakov85,Chyla89} achieves the result of renormalization scheme (RS) independent perturbative QCD,
the prize paid is that  a QCD (or Pad\'{e} QCD) coupling constant becomes defined and tied to a particular 
physical observable and its renormalization scheme  invariants $\rho_{i}.$
If therefore we observe a particular feature of such optimized QCD (or optimized Pad\'{e} QCD) 
with respect to one physical observable and its RS invariants, we cannot be sure that this feature is
universal and holds for other physical observables and their respective RS invariants.   This substitute ambiguity
applies to any finding in these optimized QCDs (or optimized Pad\'{e} QCDs), of couplant freezing as well as to 
the two scale $a_{3}, a_{4}$ infrared cut off structure discussed above. The closest we can go towards ameliorating
the ambiguity is to compute an observed couplant feature using several different QCD observables, both
timelike and spacelike, and to accept as reasonably valid and universal, only those couplant  features that
appear reasonably independent of any particular physical observable or effective charge used in computing them. To 
achieve this, we first compute here the $\rho_{2}$ values of a number of spacelike and timelike  QCD observables, whose
perturbative coefficients $r_{1}, r_{2}, r_{3}....$ are known,  at least up to NNLO loop order. We then test out
their optimized $[1|1]$ Pad\'{e} couplant structures in the infrared, and compare with our earlier findings
in paper~\cite{ndili2001} based on $R_{e+e-}(Q)$  observable.
 
 The following are the spacelike and timelike QCD observables (effective charges) we consider: 
\begin{enumerate}

\item{\bf Bjorken sum rule for deep inelastic scattering (DIS) of polarized electrons on polarized nucleons}~\cite{Bjorken66}\\
\begin{equation}\label{eq: ndili14}
\int_{0}^{1} dx \{g_{1}^{ep}(x,Q^2) - g_{1}^{en} (x, Q^2) \} =  \frac{1}{3} \left |\frac{g_{A}}{g_{V}} \right | [ 1 - R_{Bj}^{(e)} (a)]     
\end{equation}
where $R_{Bj}^{(e)}(a)$ is a QCD spacelike observable given perturbatively by:
 
\begin{equation}\label{eq: ndili15}
R_{Bj}^{(e)} (a) = a (1 + k_{1} a + k_{2} a^2 + ....... )                                              
\end{equation}
Direct loop computations in the minimal subtraction renormalization scheme $\overline{MS}$~\cite{Larin91a} give
\begin{eqnarray}
k_{1} &=& \frac{55}{12} - \frac{1}{3} N_{f}   \nonumber\\
k_{2} &=& \frac{13841}{216}  + \frac{44}{9} \zeta_{3} - \frac{55}{2} \zeta_{5} 
-  N_{f}(\frac{10339}{1296}  + \frac{61}{54} \zeta_{3} - \frac{5}{3}\zeta_{5})   \nonumber\\
      &+&  \frac{115}{648}N_{f}^2    \label{eq: ndili16}
\end{eqnarray} 

Taking $\zeta_{3} = 1.2020569031; \zeta_{5} = 1.0369277551$, and combining with 
eqns.~(\ref{eq: ndili7}),~(\ref{eq: ndili8}),~(\ref{eq: ndili9}) and~(\ref{eq: ndili13a}) we compute the
values of $\rho_{2}^{Bj(e)}$ shown in  Table~\ref{tab: ndili1}.

\item{\bf Bjorken sum rule for DIS of neutrinos on nucleons}~\cite{Bjorken66}\\ 
\begin{equation}\label{eq: ndili17}
\int_{0}^{1} dx \{F_{1}^{\bar \nu p}(x,Q^2) - F_{1}^{\nu p}(x,Q^2) \} = 1 - \frac{C_{f}}{2} (R_{Bj}^{(\nu)})    
\end{equation}

where $R_{Bj}^{(\nu)}(a)$ is a QCD spacelike observable given perturbatively by:
 
\begin{equation}\label{eq: ndili18}
R_{Bj}^{(\nu)} (a) = a (1 + f_{1} a + f_{2} a^2 + ....... )                                              
\end{equation}

Direct loop computations in $\overline{MS}$~\cite{Larin91b} give:
\begin{eqnarray}
f_{1} &=& \frac{3}{2}\left (\frac{23}{6} - \frac{8}{27} N_{f} \right )   \nonumber\\
f_{2} &=& \frac{3}{2} \left [\frac{4075}{108}  - \frac{622}{27} \zeta_{3} + \frac{680}{27} \zeta_{5} 
-  N_{f}(\frac{3565}{648}  - \frac{59}{27} \zeta_{3} + \frac{10}{3}\zeta_{5})  \right.  \nonumber\\ 
     &+&  \left. \frac{155}{972}N_{f}^2 \right ]   \label{eq: ndili19}
\end{eqnarray} 
Again one computes the corresponding $\rho_{2}^{Bj(\nu)}$ values shown in Table~\ref{tab: ndili1}, from
eqn.~(\ref{eq: ndili13a}).
 
\item{\bf Gross - Llewellyn Smith (GLS) sum rule}~\cite{Gross69}\\
\begin{equation}\label{eq: ndili20}
\int_{0}^{1} dx \{F_{3}^{\bar \nu p}(x,Q^2) + F_{3}^{\nu p}(x,Q^2) \} = 6[1 - R_{GLS}^{(\nu)}]    
\end{equation}

where $R_{GLS}^{(\nu)}(a)$ is a QCD spacelike observable given perturbatively by:
 
\begin{equation}\label{eq: ndili21}
R_{GLS}^{(\nu)} (a) = a (1 + s_{1} a + s_{2} a^2 + ....... )                                              
\end{equation}

Direct loop computations in $\overline{MS}$~\cite{Larin91a,Chyla92} give:
  
\begin{eqnarray}
s_{1} &=& \frac{55}{12} - \frac{1}{3} N_{f}   \nonumber\\
s_{2} &=& \frac{13841}{216}  + \frac{44}{9} \zeta_{3} - \frac{55}{2} \zeta_{5} 
-  N_{f}(\frac{10009}{1296}  + \frac{91}{54} \zeta_{3} - \frac{5}{3}\zeta_{5}) 
+ \frac{115}{648}N_{f}^2  \nonumber\\  
     &=& 41.441 - 8.02N_{f}  + 0.177N_{f}^2 \label{eq: ndili22}
\end{eqnarray} 

The corresponding  $\rho_{2}^{GLS(\nu)}$ values we computed from eqn.~(\ref{eq: ndili13a}) are shown in 
Table~\ref{tab: ndili1}.

\item{\bf Total hadronic cross section $R(s)$  in $e^+e^-$ annihilations}\\
\begin{equation}\label{eq: ndili23}
R(s) =  3 \sum_{f} Q_{f}^2( 1 + R_{e^+e^-}(a))
\end{equation} 
where $R_{e^+e^-}(a)$ is a QCD timelike observable given perturbatively by:
 
\begin{equation}\label{eq: ndili24}
R_{e^+e^-} (a) = a (1 + r_{1} a + r_{2} a^2 + ....... )                                              
\end{equation}
with coefficient values given in $\overline{MS}$ given  by~\cite{Gorishny91a,Surguladze91,Chetyrkin79,Dine79,Celmaster80} 
\begin{eqnarray}
r_{1} &=& 1.9857 - 0.1153 N_{f}   \nonumber\\
r_{2} &=& - 6.6368 - 1.2001 N_{f} - 0.0052 N_{f}^2 - 1.2395 \frac{(\sum Q_{f})^2}{3 \sum Q_{f}^2} \label{eq: ndili25}
\end{eqnarray} 
The corresponding $\rho_{2} = \rho_{2}^{(e^+e^-)}$ values were already computed from eqn.~(\ref{eq: ndili13})
and used in our paper~\cite{ndili2001}. They are reproduced here in Table~\ref{tab: ndili1} for direct
comparison. Formula eqn.~(\ref{eq: ndili13}) tends to make the $\rho_{2}$ values larger and more negative at the 
higher flavors $N_{f}$, than eqn.~(\ref{eq: ndili13a}).

\item{\bf $\tau$ Lepton hadronic  decay ratio $R_{\tau}$}~\cite{Chyla91,Samuel91,Braaten92}\\
\begin{equation}\label{eq: ndili26}
R_{\tau} = \frac{\Gamma (\tau \rightarrow \nu_{\tau} + hadrons)}{\Gamma (\tau \rightarrow \nu_{\tau} e^- \bar \nu_{e})} 
\end{equation}

This has a timelike QCD component observable $R_{\tau}^{\mathrm{pert}}$ given perturbatively by~\cite{Braaten92}:
 
\begin{eqnarray}
R_{\tau}^{\mathrm{pert}} &=& 3(|V_{ud}|^2 + |V_{us}|^2) \left [ 1 + a + (F_{3} - \frac{19}{24} F_{1}) a^2 \right.  \nonumber\\
                         &+&  \left. (F_{4} - \frac{19}{12} F_{3} F_{1} - \frac{19}{24} F_{2} 
+ \frac{265}{288} F_{1}^2) a^3 \right ]    \label{eq: ndili27}                                            
\end{eqnarray}
where  
\begin{eqnarray}
F_{1} &=& (2N_{f} - 33)/6  =  - b  \nonumber\\
F_{2} &=&  (19N_{f} - 153)/12  \nonumber\\
 F_{3}     &=& 1.9857 - 0.1153 N_{f}   \nonumber\\
 F_{4}     &=& -6.6368 - 1.2001 N_{f} - 0.0052 N_{f}^2    \label{eq: ndili28}
\end{eqnarray} 
and $V_{ud}$ and  $V_{us}$ are usual flavor mixing (CKM) parameters we do not however need here. The 
$\rho_{2}^{(\tau)}$ values computed  from eqn.~(\ref{eq: ndili13a}) are shown in Table~\ref{tab: ndili1}.

\item{\bf Higgs hadronic Decay Width}~\cite{Gorishny91b,Chetyrkin97}\\
\begin{equation}\label{eq: ndili29}
\Gamma_{\mathrm{tot}}(H \rightarrow hadrons)  = \frac{G_{F} M_{H}}{4 \surd 2 \pi} \sum_{N_{f}} m_{N_{f}}^2 R^{S}(s)
\end{equation}

where $R^{S}(s)$ is a QCD timelike observable given  perturbatively  up to NNLO  by~\cite{Chetyrkin97}:
\begin{eqnarray}
R^{S}(s) &=& 3 \left \{ 1 + 5.66667 a + (35.93996 - 1.35865 N_{f}) a^2  \right.   \nonumber\\
         &+&  \left.  (164.13921 -25.77119N_{f} + 0.258974 N_{f}^2) a^3 \right \}   \label{eq: ndili30}
\end{eqnarray}
or writing \\

$R^{S}(s) = 3(1 + R(s,a))$ with $R(s,a) =   a (1 + h_{1} a + h_{2} a^2 + ....... )$\\                                              

we have :  
\begin{eqnarray}
h_{1} &=&  \frac{35.93996 - 1.35865 N_{f}}{5.66667}   \nonumber\\
h_{2} &=&  \frac{(164.13921 - 25.77119 N_{f} + 0.258974 N_{f}^2)}{5.66667}  \label{eq: ndili31}
\end{eqnarray} 
The corresponding $\rho_{2} = \rho_{2}^{H}$ values computed from eqn.~(\ref{eq: ndili13a}) are shown 
in Table~\ref{tab: ndili1}.
\end{enumerate}

\section{PAD\'{E} COUPLANT FEATURES FOUND WITH SPACELIKE AND TIMELIKE OBSERVABLES}
\begin{figure}
\scalebox{1.0}{\includegraphics{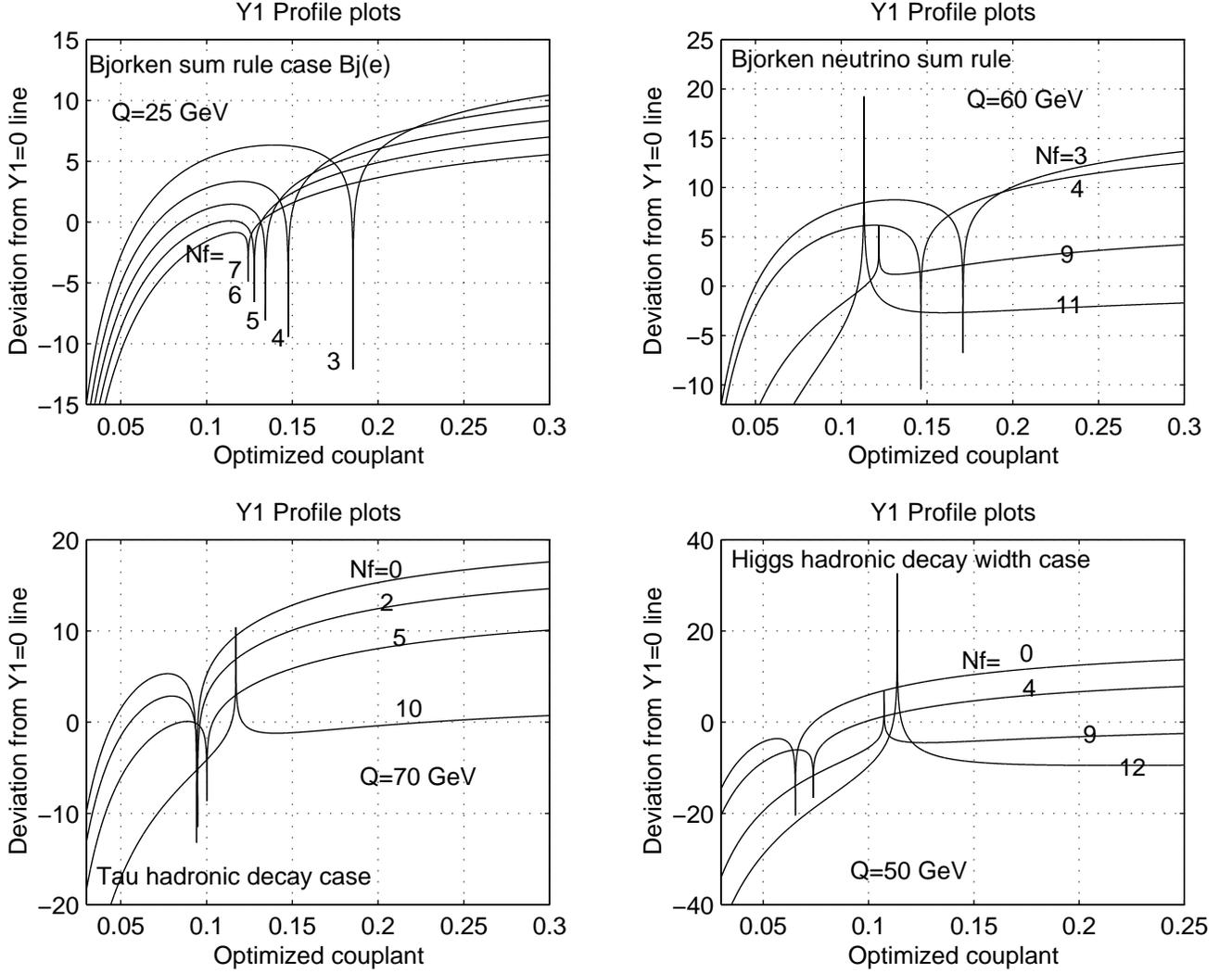}}
\caption{A plot showing similar $Y_{1}$ triple point crossing feature or solutions $a_{1}, a_{2}, a_{3}$ for
various physical observables, spacelike  and timelike, exactly as for the $R_{e+e-}(Q)$ observable discussed 
in paper~\cite{ndili2001}.}
\label{fig: ndili601}
\centering
\end{figure}

\begin{figure}
\scalebox{1.0}{\includegraphics{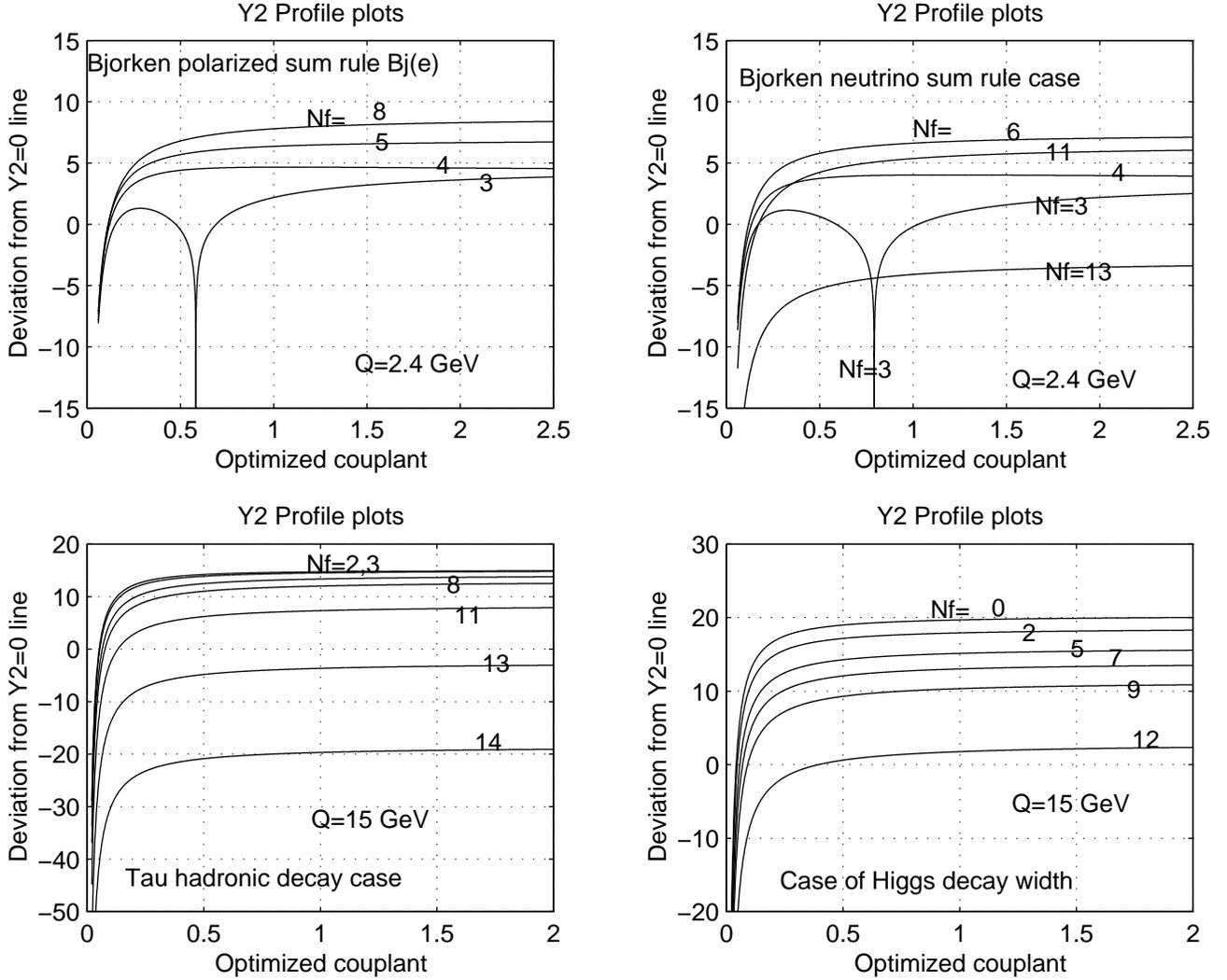}}
\caption{A plot showing similar one crossing point feature or solution $a_{4}$ of $Y_{2}$ for various QCD observables,
spacelike and timelike, except for $N_{f} \le 3$ in the spacelike cases, when $Y_{2}$ exhibits a double or  triple
crossing point feature.}
\label{fig: ndili602}
\centering
\end{figure}

\begin{figure}
\scalebox{1.0}{\includegraphics{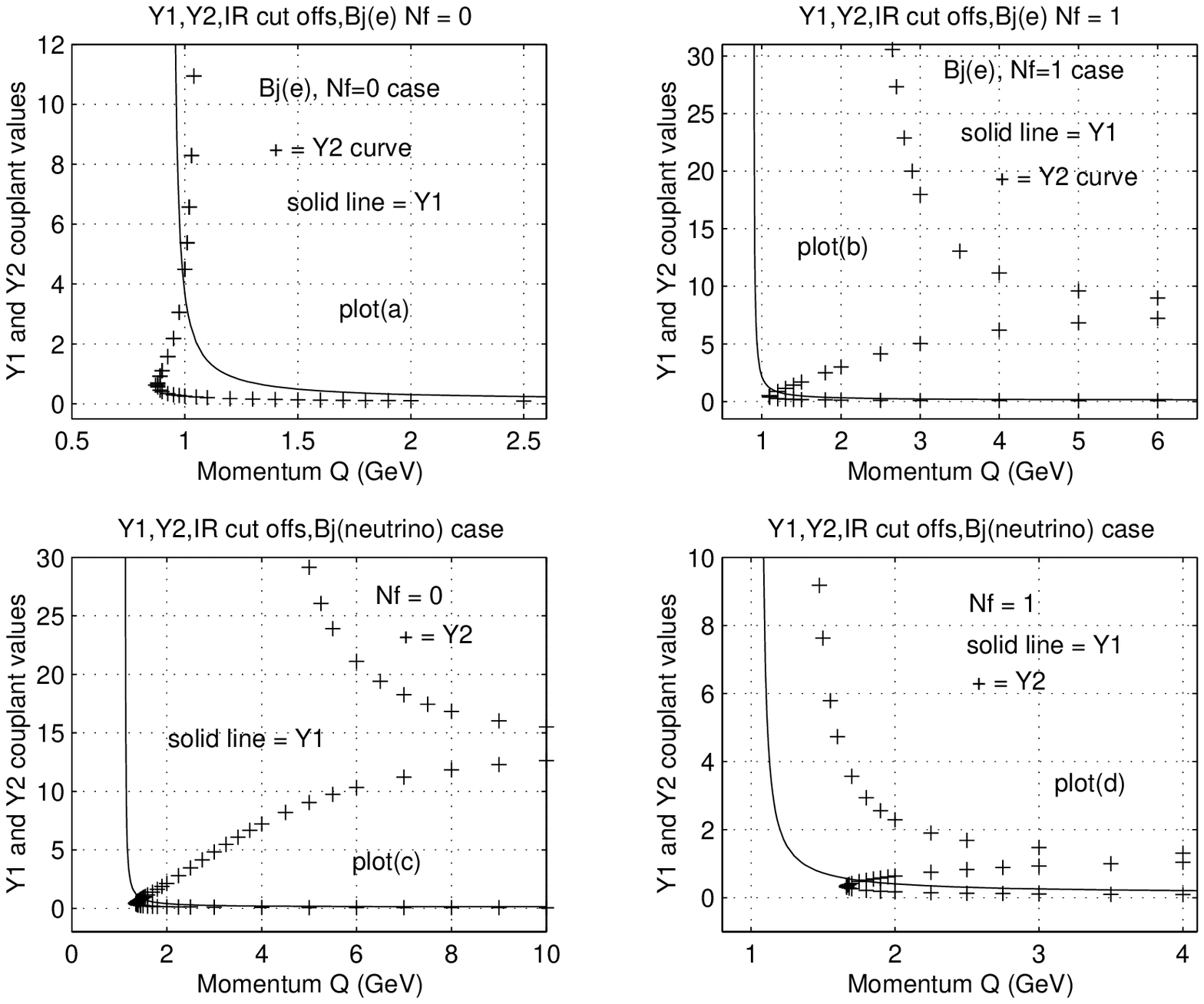}}
\caption{The  cases of $N_{f} = 0, 1$, infrared cut off structure of  $Y_{1}$ and $Y_{2}$  for the two spacelike  
 observables, Bj(e) and Bj($\nu$) for which $\rho_{2}$  is positive and large, and $D$ is  negative}.
\label{fig: ndili608}
\centering
\end{figure}

\begin{figure}
\scalebox{1.0}{\includegraphics{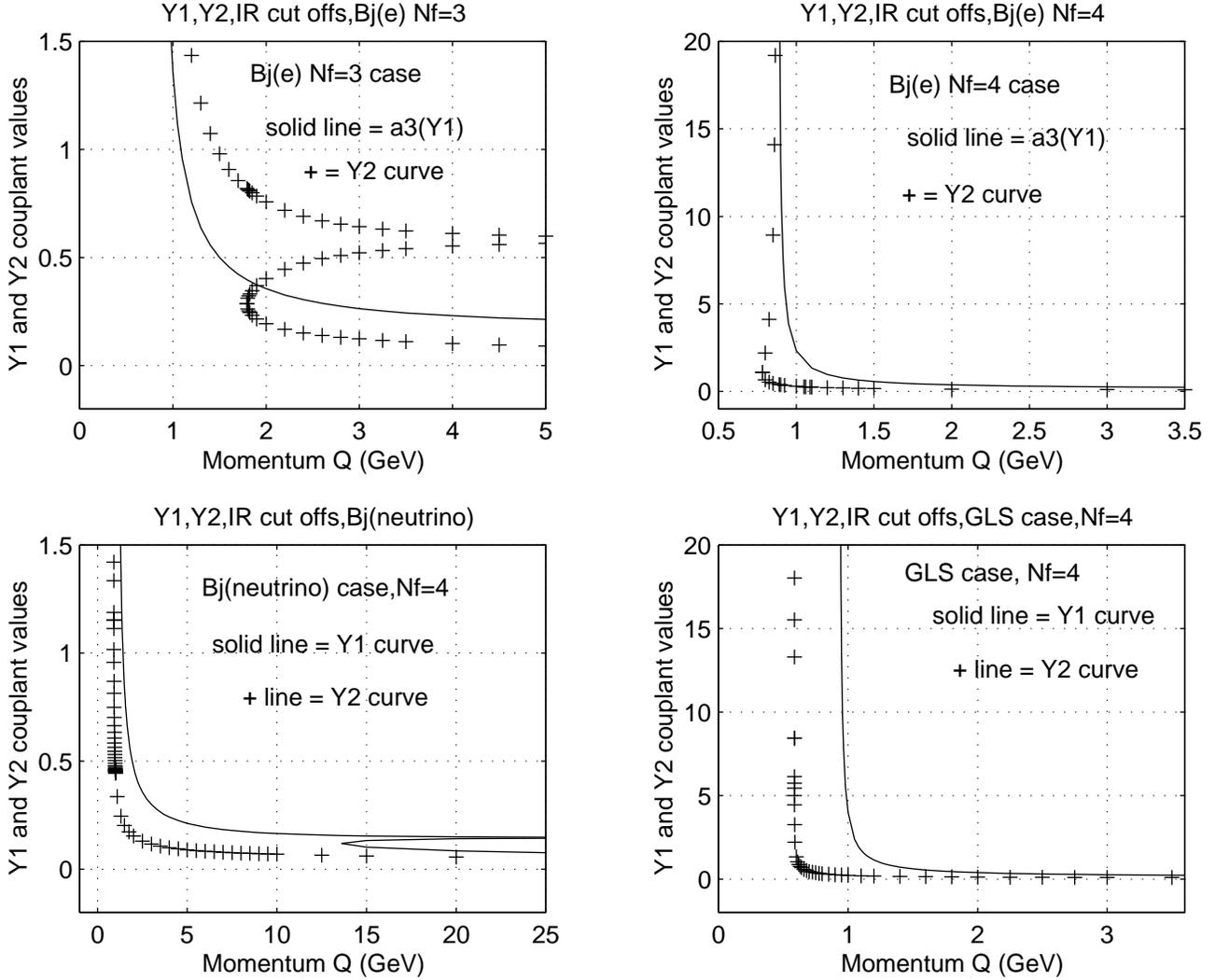}}
\caption{A plot showing the  $Y_{2}$ cut off behavior in the $N_{f} = 3, 4$ flavor states of 
positive $\rho_{2}$ but positive $D$, in spacelike  physical observables, Bj(e), Bj($\nu$), and
Gross - Llewellyn Smith (GLS) sum rules.  The cut offs $a_{3}, a_{4}$ are seen
to occur closely together, with $a_{4}$ even growing faster than  $a_{3}$ in the ensuing CQCD dynamics
in $N_{f} = 3$ case}.
\label{fig: ndili609}
\centering
\end{figure}

Given now the above  $\rho_{2}$ values for different observables shown in Table~\ref{tab: ndili1}, we use the same
computational procedures described in  paper~\cite{ndili2001}, to ascertain the multicomponent couplant structures
of eqns.~(\ref{eq: ndili1}) and~(\ref{eq: ndili2}) with respect to each of the above  observables and  their $\rho_{2}$ 
values. Our findings and  comparative features are  as follows:
  
\begin{enumerate}
  
\item First we note from Table~\ref{tab: ndili1} that the spacelike observables Bj(e), Bj($\nu$), GLS($\nu$)
have positive $\rho_{2}$ values for $N_{f} \le 4$, except for a small negative $\rho_{2} = -0.3223$ at
$N_{f} = 4$ in the GLS case.  Correspondingly, we find that the quantity $D$ of eqn.~(\ref{eq: ndili6}) 
is negative for $N_{f} = 0,1,2$ in these spacelike observables.  In these negative
$D$ cases, eqn.~(\ref{eq: ndili54}) has  no real $\bar c_{2}$ solution, and progress is possible only if we use
$\sqrt{|D|}$ in eqns.~(\ref{eq: ndili4}) and~(\ref{eq: ndili5}).  For $N_{f} = 3, 4$, the quantity $D$ is
positive even though $\rho_{2}$ remains positive.  Subject to these observations, 
equations~(\ref{eq: ndili4}) and~(\ref{eq: ndili5}) apply exactly as in the timelike cases, and our
observed couplant features should become comparable in all spacelike and timelike cases. 
  
\item Computing the crossing point features of eqn.~(\ref{eq: ndili1}) for $N_{f} \geq 3,$
we find exactly the same triple point $Y_{1} = a_{1}, a_{2}, a_{3}$ solution for the spacelike observables as for
the timelike observables. This is shown in fig.~\ref{fig: ndili601} and agrees with the features  we found in 
paper~\cite{ndili2001}. The same  triple crossing point solutions $Y_{1} = a_{1}, a_{2}, a_{3}$ 
are also found for  $N_{f} = 0, 1, 2$ flavor states of the spacelike observables, provided we  use $|D|$ in
eq.~(\ref{eq: ndili1}).  They all give similar $Y_{1}$ spiral structures with parameter values
shown in Tables~\ref{tab: ndili2},~\ref{tab: ndili3}, and~\ref{tab: ndili4}, to be
compared with similar structural parameters given in paper~\cite{ndili2001} for the $R_{e+e-}(Q)$ observable.

\item  When we compute next the crossing point features of $Y_{2}$ from eqn.~(\ref{eq: ndili2}), we find that 
in general for $N_{f} \le 4$ the spacelike observables behave differently from the timelike observables.  The $Y_{2}$ 
plots of these spacelike observables for $N_{f} \le 3$, show  a triple point  crossing  feature: 
$Y_{2} = b_{1}, b_{2}, b_{3},$  much like their $Y_{1} = a_{1}, a_{2}, a_{3}$ counterpart.  The $N_{f} = 3$ case of this
$Y_{2}$ triple crossing feature can be seen in  fig.~\ref{fig: ndili602}.  In contrast, for $N_{f} \geq 5$, all physical 
observables, spacelike and timelike, show only one $Y_{2}$  crossing point solution $a_{4}$ seen  in 
fig.~\ref{fig: ndili602} and comparable to our earlier findings in paper~\cite{ndili2001}.  The above triple $Y_{2}$ 
crossing behavior for $N_{f} \le 3$ in the spacelike cases, will be found later  to lead to a peculiar infrared cut
off behavior of $Y_{2}$ relative to $Y_{1}$ in these low flavor states of spacelike observables.

\item We look closely next at the infrared cut off region of these  spacelike and timelike observables. 
Taking first the cases of $N_{f} = 5, 6 ...$ for  the spacelike observables and looking in
their infrared cut off region shown in figs.~\ref{fig: ndili605} and~\ref{fig: ndili605a}, we see exactly 
the same two scale cut off structure, as that shown in figs.~\ref{fig: ndili600a} and~\ref{fig: ndili600b}
for the timelike $R_{e+e-}(Q)$ case.  The structures separate QCD dynamics in momentum transfer (or energy)
space into three distinct regimes we have labelled in the figures as PQCD, CQCD, and hadron QCD.

\item  Exactly this same two  distinct scale cut off structure is  exhibited for all flavor states 
$0 \le N_{f} \le  8$ by the $\tau$  hadronic decay rate and the Higgs hadronic decay  width observables shown in part 
in figs.~\ref{fig: ndili606} and ~\ref{fig: ndili607}.  The figures compare well with 
figs.~\ref{fig: ndili600a} and~\ref{fig: ndili600b} of the $R_{e+e-}(Q)$ observable.

\item  The peculiar cases of $N_{f} \le 4,$ for spacelike observables are shown in figs.~\ref{fig: ndili600c} 
to~\ref{fig: ndili609}, to be compared with the same flavor state plots of timelike observables 
shown variously in figs.~\ref{fig: ndili600a},~\ref{fig: ndili600b},~\ref{fig: ndili606}, and~\ref{fig: ndili607}.  
We see that arising directly from the positive $\rho_{2}$ values in these $N_{f} \le 4,$ flavor states of spacelike 
observables, their $Y_{2}$ and $Y_{1}$ curves tend to bunch together at the infrared cut off, with the
$Y_{2}$ appearing to cut off earlier than the $Y_{1}$ curve for $N_{f} \le 3$, though not for $N_{f} = 4$.
The bunching together tends to reduce the CQCD momentum gap or domain, compared to figs.~\ref{fig: ndili600a}
and~\ref{fig: ndili600b}, or figs.~\ref{fig: ndili605} to~\ref{fig: ndili607}.  However, the two scale infrared cut 
off structure still remains a finding, even in these positive $\rho_{2}$ spacelike cases.

\item Having seen that the two scale nonperturbative infrared structure persists in all flavor states $N_{f} \le 8$,
and for all observables spacelike and timelike, we now consider the paired numerical values of these scales,
$Q_{c}(Y_{2})$ and $Q_{c}(Y_{1})$, and how these values vary from flavor to flavor and from one QCD  observable (or
perturbative series) to another.  There are two obvious points at which we can evaluate and compare the two scales. 
We can evaluate the two scales at the point where each curve $Y_{2}$ and $Y_{1}$ independently begins to depart 
appreciably from its smoothly rising asymptotically free  PQCD curve, to enter a nonperturbative QCD phase characterised by
sudden rapidly rising  coupling constant $a_{4}$ and $a_{3}$ respectively, in the infrared region.  We shall denote the
values of $Q_{c}(Y_{2})$ and $Q_{c}(Y_{1})$ so evaluated at this onset point of nonperturbative infrared QCD dynamics, by
$Q_{on}(Y_{2})$ and  $Q_{on}(Y_{1})$ respectively.  The second obvious point where we can evaluate and compare the two
scales is where each curve independently cuts off at the far infrared end, signalled by its coupling constant rising
to infinite values $(a \geq 30$, or $\alpha_{s} \geq  100)$.  We shall denote the two scales evaluted at this terminal 
infrared cut off point for each curve, by $Q_{c}(Y_{2}) =  Q_{off}(Y_{2})$ and $Q_{c}(Y_{1}) =  Q_{off}(Y_{1})$ respectively. 
A third obvious  evaluation point exists in the special cases of $N_{f} \le 4$  for spacelike observables, where as shown  
in  figs.~\ref{fig: ndili600c} to~\ref{fig: ndili609}, the $Y_{2}$ curve exhibits a sudden turning away from the infrared
region within the nonperturbative QCD regime, before finally rising up to infinity. This $Y_{2}$ infrared attractor point, 
provides a distinct coupling constant hierachy  point at which we can also evaluate and compare the two infrared
nonperturbative $Y_{2}$ and  $Y_{1}$ scales.

\item Regarding the onset point of nonperturbative dynamics,   we find from  our various plots shown in
figs.~\ref{fig: ndili600a},~\ref{fig: ndili600b}, and~\ref{fig: ndili605} to~\ref{fig: ndili609}, that the two curves 
$Y_{2}$ and $Y_{1}$ commence their transition into  the nonperturbative regime at about the same critical
coupling constant value, seen from the plots to lie in the range  $0.18 \le a_{\mathrm{con}} \approx a_{\chi} \le 0.28$, 
or $0.56 \le \alpha_{s}^{\mathrm{con}} \approx \alpha_{s}^{\chi} \le  0.87$, where $a_{\mathrm{con}}$ denotes this
critical coupling as evaluated on the $Y_{2}$ curve, while  $a_{\chi}$ denotes the same nonperturbative  onset  critical
coupling as evaluated on the $Y_{1}$ curves.  If we look at the $Y_{2}$ plots as the more sharply focused indicator 
of this point of coupling constant criticality, we find from figs.~\ref{fig: ndili600b},~\ref{fig: ndili606}
and~\ref{fig: ndili607}, that the $Y_{2}$ curves for different flavors  tend actually to  pass through a common point 
that lies around  $a_{\mathrm{con}} \approx  0.25$, beyond which these various flavor lines cross paths and rise sharply
to infinity. Basing on this,  we can take as the mean  onset critical coupling constant revealed by our plots  for both
$Y_{2}$ and $Y_{1}$, to be the value $a_{\mathrm{con}} \approx a_{\chi} = 0.25$,  Accordingly, we can  evaluate
and compare $Q_{on}(Y_{2})$ and  $Q_{on}(Y_{1})$ at  $a_{\mathrm{con}} \approx a_{\chi} = 0.25$.   The values of the two
nonperturbative infrared scales evaluated at this point (I) and at the other two strategic points (III, and II) stated above, 
are shown in Tables~\ref{tab: ndili5} and~\ref{tab: ndili6}.

\item  Ignoring a few  positive $\rho_{2}$ cases where specifically the $Y_{2}$ curve cut off widely earlier than the
 $Y_{1}$ curve  as shown  in  plots (b) and (c) of  fig~\ref{fig: ndili608},  we find one running feature of 
the paired scales shown in Tables~\ref{tab: ndili5} and~\ref{tab: ndili6}.  This feature is that of the two 
nonperturbative infrared cut off scales $Q_{off}(Y_{2})$ and $Q_{off}(Y_{1})$, the lower one defined by the
$Y_{2}$ solution occurs at a mean  critical momentum cut off value of $Q_{off}(Y_{2})  \approx 296$ MeV, well
within a value we would associate with $\Lambda_{\mathrm{QCD}}$, the fundamental scale or constant  of QCD.  On the other
hand, the upper cut off scale defined by $Q_{off}(Y_{1})$ occurs at a mean value: $Q_{off}(Y_{1}) = 1.44$ GeV.  Both
results are consistent with our earlier findings in paper~\cite{ndili2001} based on  timelike $R_{e+e-}(Q)$ observable.

\item  A second feature we observe readily from Tables~\ref{tab: ndili5} and~\ref{tab: ndili6}, is that irrespective
of the  point of evaluation and the particular QCD observable considered,  the two scales $Q_{c}(Y_{2})$ and
$Q_{c}(Y_{1})$  are functions of flavor number $N_{f}$, both changing in value as the flavor number runs from
$N_{f} = 0$  to $N_{f} = 8$.  Here the spacelike and timelike QCD observables exhibit a distinct difference in
their $N_{f} \le 4$ states.  Explicitly we find that for the timelike observables and for all
$0 \le N_{f} \le 8$, the two infrared scales move progressively apart as flavor increases.  This can be seen directly in 
figs.~\ref{fig: ndili600a},~\ref{fig: ndili600b},~\ref{fig: ndili606} and~\ref{fig: ndili607}.  The same feature holds
for spacelike observables but only for $N_{f} \geq 4$ shown in figs.~\ref{fig: ndili605} and~\ref{fig: ndili605a}.
For $N_{f} < 4$, Tables~\ref{tab: ndili5} and~\ref{tab: ndili6},  and figs.~\ref{fig: ndili600c} to~\ref{fig: ndili609}
show that the spacelike observables exhibit a different pattern of flavor variation. With a view to later investigating
what these patterns of flavor variation of the two nonperturbative infrared scales can  mean, we have computed and shown 
in Tables~\ref{tab: ndili5} and~\ref{tab: ndili6}, the ratio quantities $Q_{c}(Y_{2})/Q_{c}(Y_{1})$ of the two scales 
evaluated at one or the other of the three strategic  couplant points mentioned above for various flavors and QCD
observables.  The ratios have been denoted I, II, II, depending on the point of evaluation of the $Y_{2}, Y_{1}$ scales. 
Since this ratio gives a  measure of the moving apart or the  closing in of these two scales,  we shall call the ratio 
a correlation factor  between the two solutions $Y_{2}$ and $Y_{1}$ and the dynamical processes they cause or represent.

\item Summarizing, we can affirm that our curves not only define a three phase Pad\'{e} QCD regime labelled PQCD, CQCD, 
and hQCD in our figures, but give us definite critical coupling constant values and critical momenta at which one 
transits from one Pad\'{e} QCD regime into another.  The extent these features and findings from Pad\'{e} QCD lead us to 
conclude about physical QCD are what we  examine next.   

\end{enumerate}

\section{PHENOMENOLOGICAL FEATURES OF DYNAMICAL CHIRAL SYMMETRY BREAKING AND CONFINEMENT IN QCD}
\begin{figure}
\scalebox{1.0}{\includegraphics{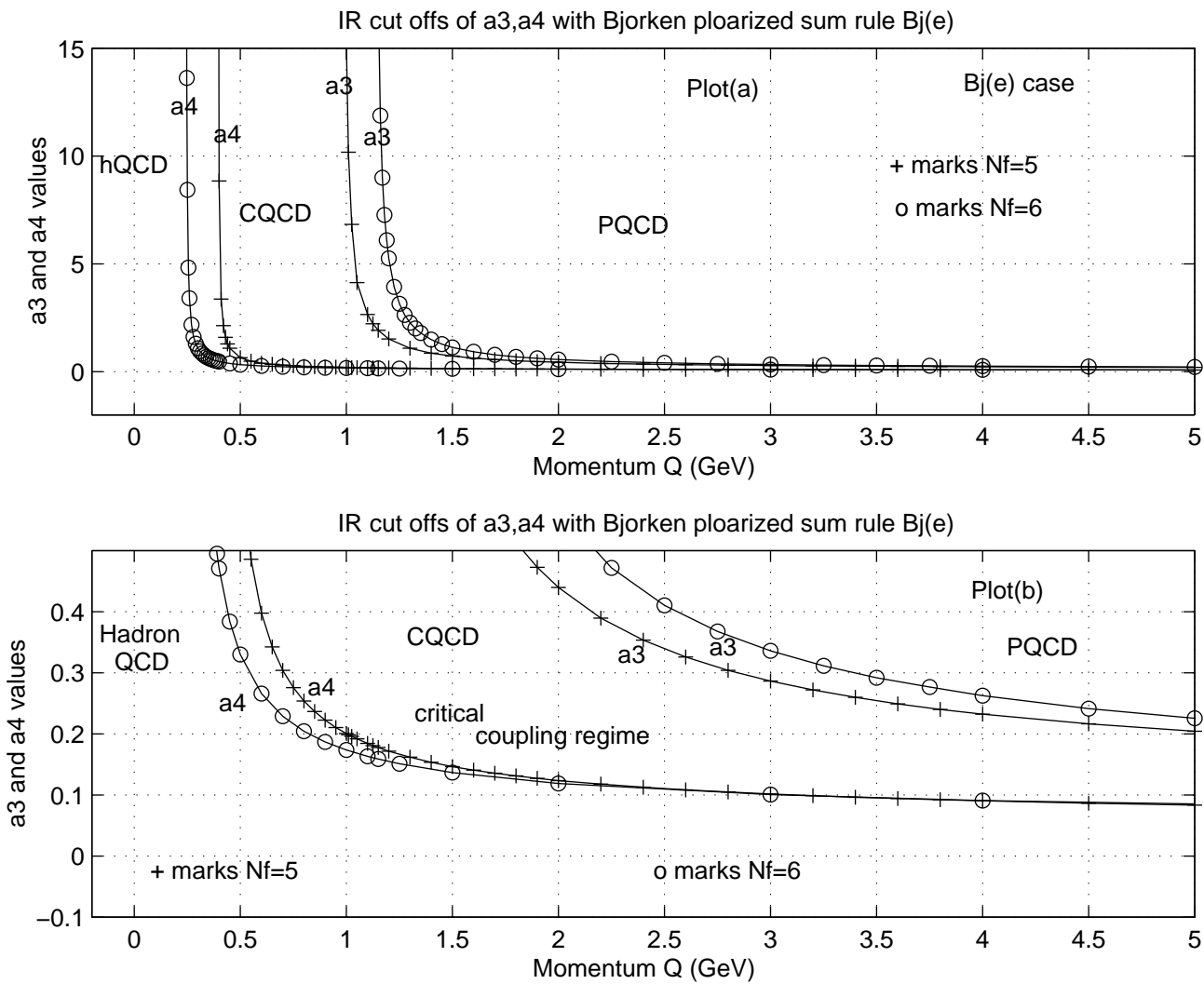}}
\caption{A two scale cut off structure of $a_{3}$ and $a_{4}$ in the infrared region:
The case of Bjorken polarized sum rule  Bj(e), (spacelike).}
\label{fig: ndili605}
\centering
\end{figure}

\begin{figure}
\scalebox{1.0}{\includegraphics{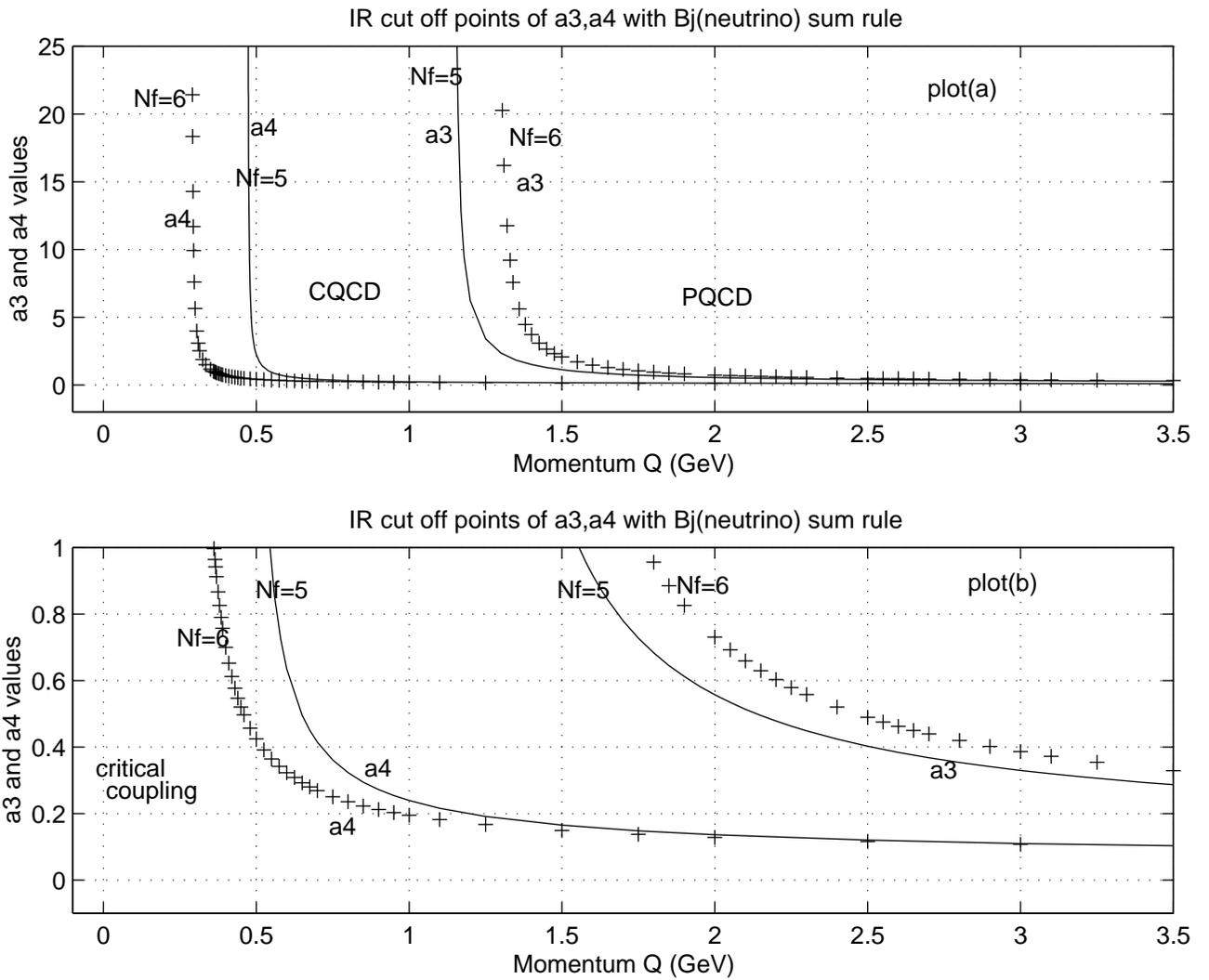}}
\caption{A two scale cut off structure of $a_{3}$ and $a_{4}$ in the infrared region:
The case of Bjorken neutrino sum rule  Bj($\nu$), (spacelike).}
\label{fig: ndili605a}
\centering
\end{figure}

\begin{figure}
\scalebox{1.0}{\includegraphics{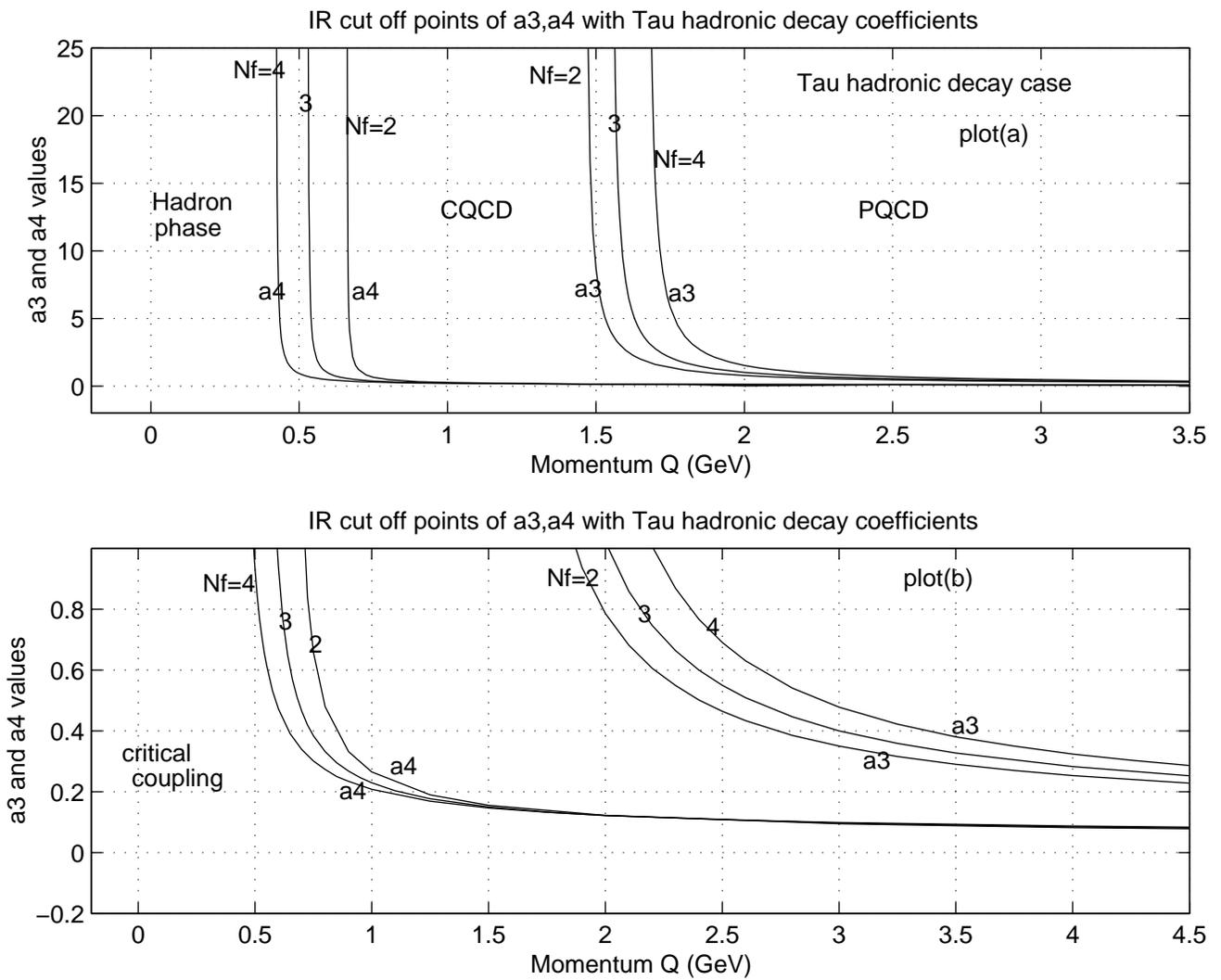}}
\caption{A two scale cut off structure of $a_{3}$ and $a_{4}$ in the infrared region:
The case of tau hadronic decay rate, (timelike).}
\label{fig: ndili606}
\centering
\end{figure}

\begin{figure}
\scalebox{1.0}{\includegraphics{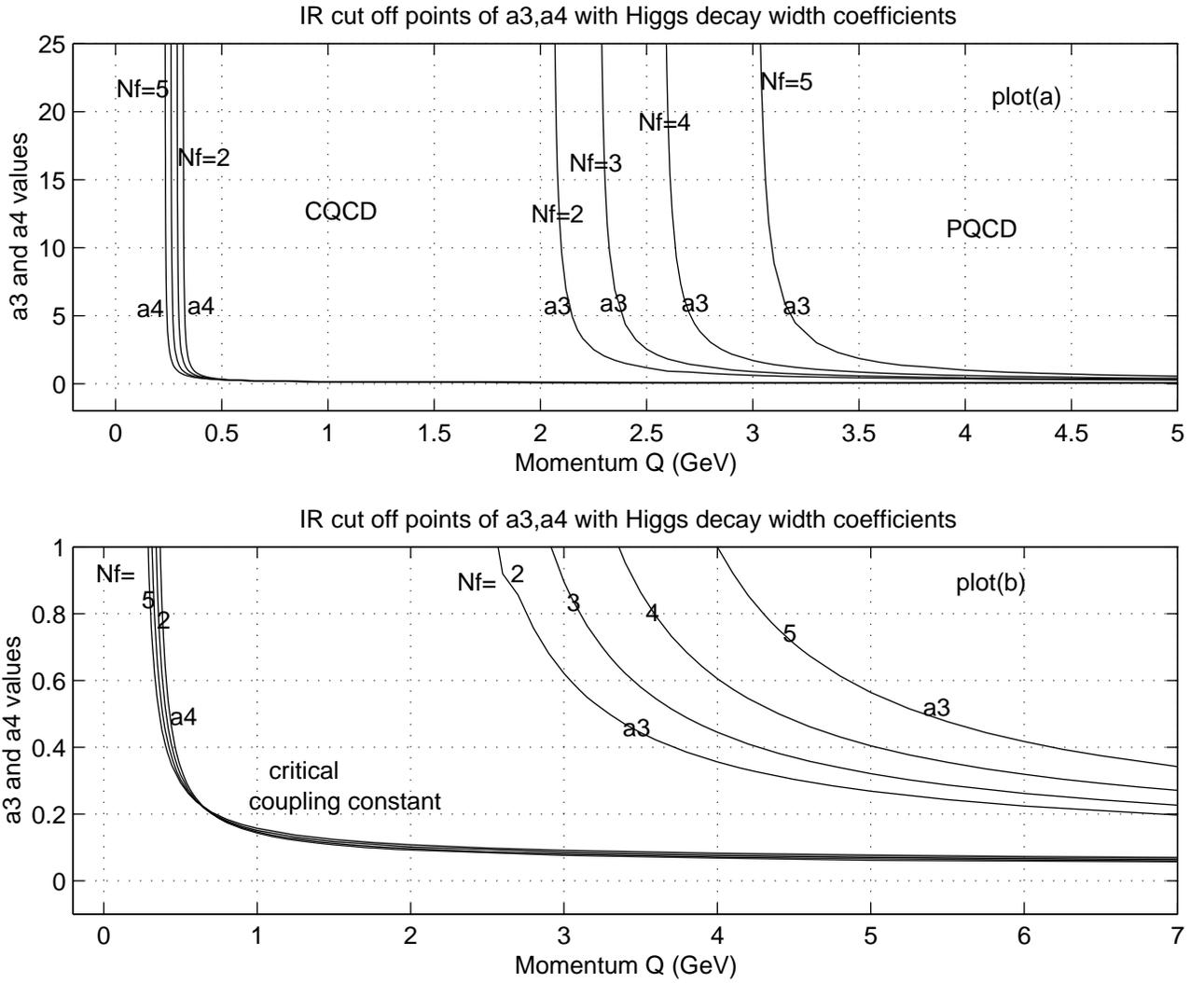}}
\caption{A two scale cut off structure of $a_{3}$ and $a_{4}$ in the infrared region:
The case of Higgs hadronic decay width, (timelike).}
\label{fig: ndili607}
\centering
\end{figure}

\begin{figure}
\scalebox{1.0}{\includegraphics{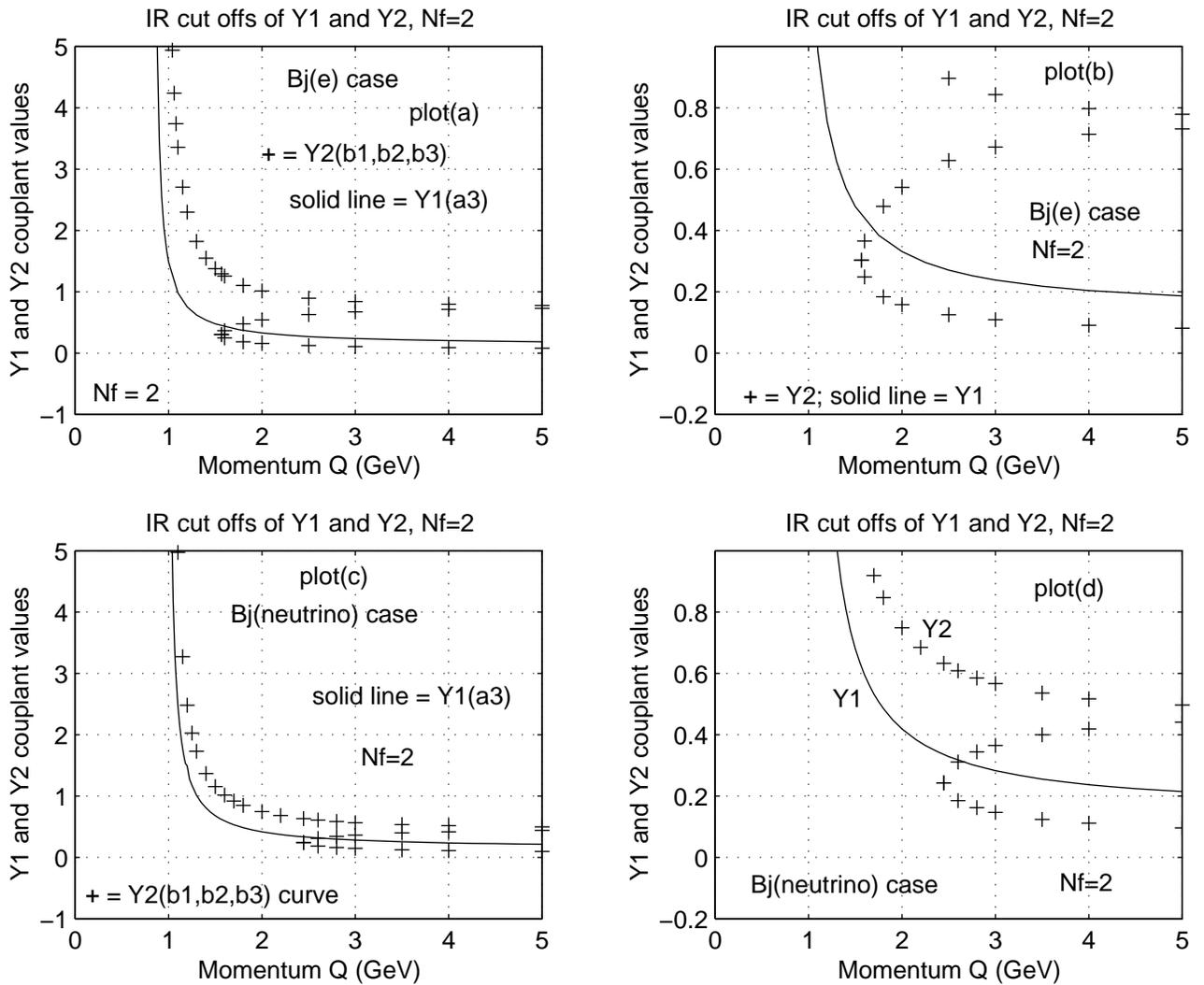}}
\caption{The  $N_{f} = 2$  infrared cut offs of  $Y_{1}$ and $Y_{2}$  plotted for two spacelike  
 observables for which  $\rho_{2}$  is positive and large, and $D$ is  negative. The cut off curves are
seen  to bunch closely together.}
\label{fig: ndili600c}
\centering
\end{figure}

To see that the above Pad\'{e} QCD infrared critical  momenta $Q_{c}(Y_{1})$  and $Q_{c}(Y_{2})$  are amenable to
interpretation as boundaries of the CQCD domain of physical QCD, we consider now some phenomenological features 
and attributes of dynamical chiral symmetry breaking $(D \chi SB)$  and  quark confinement in QCD
gleaned from a number of different sources and models, as follows:

\begin{enumerate}
\item  Miransky~\cite{Miransky93} reviewing a broad spectrum of  phenomenogical  models of low energy QCD, in particular
 Bethe Salpeter wave function models of dynamical chiral symmetry breaking $(D \chi SB)$ and quark confinement, concluded that  
the momentum  region of QCD where $D \chi SB$ takes place can be parameterized phenomenologically  as:
$\delta ^2 \le Q^2 \le \Lambda_{\chi}^2$, where Q is the QCD operating  momentum, while  $\delta$ 
and $\Lambda_{\chi}$ are phenomenological infrared (IR) and ultraviolet (UV) cut-off  momenta  respectively, 
the values of which were phenomenologically found to be:  $\delta  \geq \Lambda_{\mathrm{QCD}} \approx 300$ MeV,
and $\Lambda_{\chi} \approx 1$ GeV.  The delta scale $\delta \approx \Lambda_{\mathrm{QCD}}$ was identified with the 
confinement scale of QCD, and was thus found phenomenologically to be at least a few times smaller than the  $D \chi SB$ scale.
These values are to be compared with our findings in Tables~\ref{tab: ndili5} and~\ref{tab: ndili6} where writing
$Q_{off}(Y_{2}) \le Q_{\mathrm{CQCD}} \le Q_{off}(Y_{1})$ as $\Lambda_{\mathrm{con}} \le Q_{\mathrm{CQCD}} \le \Lambda_{\chi}$, 
we found $\Lambda_{\mathrm{con}}  =  296$ MeV (mean value), while  $\Lambda_{\chi} = 1.44$ GeV (mean value), both
for all $0 \le N_{f} \le 8$ and for various physical observables, timelike and spacelike. The agreement is good.

\item Concerning the QCD critical coupling constant at the onset of $D \chi SB$,
Refs.~\cite{Miransky93, Fomin83, Atkinson88} found from their phenomenological estimates and approximations,
the value: $\alpha_{s}^{\chi} \geq \pi /4$ meaning $a_{\chi} \geq  0.25$.  The same value was arrived  at by
Higashijima~\cite{Higashijima91} who found in his effective potential approaches, that chiral symmetry breaks down
when his parameter  $\lambda(t_{c}) = 3 C_{2}(R) g^2 /(4\pi^2)  > 1$, where $C_{2}(R) = 4/3$ for three color QCD,
implying $\alpha_{s}^{\chi} = g_{c}^2/(4 \pi) \geq \pi /4$, or $a_{\chi}  \geq  0.25$.  
These phenomenological QCD critical  couplings for $D \chi SB$ are to be compared with our exact finding in
fig.~\ref{fig: ndili607} and other plots, that the various flavor lines converge and rise nonperturbatively at 
$a_{\mathrm{con}} \approx a_{\chi} \approx 0.25$, both for timelike and spacelike observables. The agreement is 
excellent.

\item Roberts and McKellar~\cite{Roberts90} solving numerically the  Schwinger-Dyson (SD) 
integral equation for quark self energy (condensate), under varying kernel approximations, obtained the specific values:
$\alpha_{s}^{\chi} = 0.782$  in one kernel approximaion, and  $\alpha_{s}^{\chi} = 0.890$ in a more exact  kernel
evaluation.  Their results then place $a_{\chi}$ in the range   $0.253  < a_{\chi} \le 0.29$  for  $N_{f} = 0, 4$
considered by them, to be compared again with our above Pad\'{e} QCD finding in the general range: 
$0.18 \le a_{\chi} \le  0.28$ for $0 \le N_{f} \le 6$.  The agreement is again excellent.  Roberts and McKellar 
additionally found from their work that $D \chi SB$ and confinement do not necessarily occur together; rather
the latter is held to occur subsequently, close to where the QCD coupling constant grows infinitely large as momentum 
transfer goes to zero. Our figs.~\ref{fig: ndili600} to~\ref{fig: ndili600b}, and figs.~\ref{fig: ndili605} 
to~\ref{fig: ndili609} manifest exactly this two scale feature and a sharply rising coupling constant to infinity,
compatible with infrared slavery confinement at very low Q values.

\item  Earlier work by Atkinson and Johnson~\cite{Atkinson87} established the interesting point that  there exists a
critical value of QCD coupling constant, corresponding to the onset of chiral symmetry breaking, provided that (a) there 
is an infrared cutoff, realized phenomenologically as an effective gluon mass, and (b) there is an ultraviolet cutoff
linked with a running  coupling constant (a chiral symmetric phase of QCD). Within their own approximations to a
fermion propagator Schwinger-Dyson equation, they find the  critical QCD  coupling constant for $D \chi SB$  to have the value
$\alpha_{s}^{\chi} = 0.91$.   These findings from the Schwinger-Dyson ladder approximant equation agree again 
quite well with our Pad\'{e} QCD two scale infrared NPQCD structure, and with  the fact that our plots show
$\alpha_{s}^{\chi} \approx 0.87$  at  the upper limit.

\item  In more recent studies, Guo and Huang~\cite{Guo97} using two different approaches: (a) a semi-phenomenological
self consistent equation to determine the quark condensate (the order parameter of $D \chi SB$) in the chiral limit,
and (b) the usual Schwinger-Dyson  equation (SD) improved by addition of gluon condensate kernel, obtained a critical QCD
coupling $\alpha_{s}^{\chi}$ above which chiral symmetry breaks down.  Their value is: $\alpha_{s}^{\chi} = 0.2 \pi$,
which means $a_{\chi} = \alpha_{s}^{\chi}/ \pi = 0.2$.  This again agrees very well with our Pad\'{e} QCD findings seen
in fig.~\ref{fig: ndili600} and figs.~\ref{fig: ndili605a} to~\ref{fig: ndili600c}, and  our specific mean value:
$a_{\mathrm{con}} \approx a_{\chi} = 0.25$.

\item Approaching our CQCD regime of QCD from the hadron end,  in heavy ion Quark Gluon Plasma (QGP) studies, Lattice
gauge theories and simulations~\cite{Baal98, Kogut82, Kogut83}, have evidence of the same two scale NPQCD structure found from
our Pad\'{e} QCD, and in the correct sequence. They find that the chiral symmetry restoration temperature $T_{\chi}$ is 
somewhat larger than the deconfinement temperature $T_{\mathrm{decon}}$.  Put differently, Lattice QCD studies find
that $D \chi SB$ is characterized by smaller distances compared to confinement distances.
In our Pad\'{e} QCD results played backwards, we expect our lower cut off momentum
$Q_{off}(Y_{2}) = \Lambda_{\mathrm{con}}$ to identify closely with $T_{\mathrm{decon}}$ as two parameters that mark 
confinement/deconfinement in QCD.   Correspondingly, we expect  our upper cut off momentum:
$Q_{off}(Y_{1}) = \Lambda_{\chi}$ to also identify closely with $T_{\chi}$ as two parameters that mark chiral symmetry 
breaking and restoration in QCD dynamics.  Our Pad\'{e} QCD finding  that in general
$\Lambda_{\chi} > \Lambda_{\mathrm{con}}$  then receives good support from the  above independent Lattice QCD studies.

\item Explicitly, Shuryak~\cite{Shuryak81} argued from a variety of phenomenological observations, including power
corrections to deep inelastic scattering and  heavy ion quark gluon plasma processes, that  there must exist in the 
infrared region of QCD, a second scale other than the confinement scale, and that this second scale must be connected with chiral
symmetry breaking.  He estimated the hierachy of the two scales to be:
$\Lambda_{\mathrm{\chi SB}} \gg  \Lambda_{\mathrm{con}} \approx  \Lambda_{\mathrm{QCD}}$, and that their ratio is of
the order 3 - 5, meaning reciprocal ratios 0.33 - 0.20.  Such two scale structure and hierachy  in the infrared region
of QCD is what we find here from our Pad\'{e} perturbative QCD analysis shown variously in
figs.~\ref{fig: ndili600} to~\ref{fig: ndili600b} and figs.~\ref{fig: ndili605} to~\ref{fig: ndili609}.  
Our $Q_{c}(Y_{2})/Q_{c}(Y_{1})$ ratios shown in Tables~\ref{tab: ndili5} and~\ref{tab: ndili6}  for our two 
infrared scales are well within the estimates of Shuryak, depending on flavor.

\item In the context of effective Lagrangian theories of the strongly interacting regime of QCD where chiral symmetry 
breaking and quark confinement are believed to occur, Nambu and Jona-Lasinio (NJL)~\cite{Nambu61} were the first to 
introduce a viable effective Lagrangian model of hypothetical four fermion interactions, in which chiral symmetry breaking 
with pions as Goldstone bosons, was fully realized. The QCD scale at which the model worked entered the NJL effective
Lagrangian theory as an upper cut-off momentum and  was explicitly found to be 
$\Lambda_{\mathrm{\chi SB}} \approx 1$ GeV.  Additionally, Nambu and Jona-Lasinio found that their effective theory works 
in a gap regime of QCD lying between the scale $\Lambda_{\mathrm{\chi SB}} \approx 1$ GeV and some lower hadronic
scale $(\approx 300 MeV)$ believed to be the onset point of quark confinement  into hadrons.  What we now observe is that
this NJL model  chiral symmetry breaking scale $\Lambda_{\mathrm{\chi SB}} \approx 1$ GeV, as well as the gap regime
stretching to lower confinement scale, are exactly comparable with our finding from our Pad\'{e} plots that, 
$296 MeV  \le  Q_{\mathrm{CQCD}}  \le  1.44 GeV$, the implication of all this being that we  identify  our Pad\'{e} 
CQCD domain with the NJL gap regime  of Goldstone physics. 

\item Another striking phenomenological model buttressing our Pad\'{e} case is that of Manohar and 
Georgi~\cite{Manohar84} who not only found the phenomenological necessity for the two scale
$(\Lambda_{\mathrm{\chi SB}}, \Lambda_{\mathrm{con}})$ Goldstone physics structure and an intermediate infrared QCD
regime bounded by $\Lambda_{\mathrm{con}} < Q < \Lambda_{\mathrm{\chi SB}}$ with
$\Lambda_{\mathrm{\chi SB}} \gg \Lambda_{\mathrm{con}} \approx \Lambda_{\mathrm{QCD}}$, but actually proposed the use
of the ratio of the two scales as a natural expansion parameter in chiral effective Lagrangian theories of low energy QCD.
Their two scale  structure has numerical values $\Lambda_{\mathrm{\chi SB}} \approx 1$ GeV, and
$\Lambda_{\mathrm{con}} \approx 100 - 300$ MeV, both very close to our Pad\'{e} QCD findings.

\item In their own novel   gauge invariant nonperturbative approach to QCD at large distances, Gogohia and 
co-workers~\cite{Gogohia90} found that a momentum free parameter $k_{o}$ in their approach, defined as some
characteristic scale at which confinement and other nonperturbative ($D \chi SB)$ effects begin to play 
a dominant role in QCD,  or more broadly defined as a momentum that separates the nonperturbative phase of QCD from the
perturbative phase, always has both a lower and  an upper boundary value which  Gogohia et. al. found numerically 
to be: $635 MeV  \le k_{o} \le 775 MeV$. Their more recent estimates~\cite{Gogohia94,Gogohia99} give a higher value
$\Lambda_{\mathrm{\chi}} \approx 1.2$ GeV, depending on  input value for the pion decay constant.  These result can be 
interpreted as a finding of a two scale  structure in long distance QCD  dynamics, and  the explicit scale  values 
estimated above by Gogohia et. al. are not incomparable with our paired scales shown in 
Tables~\ref{tab: ndili5} and~\ref{tab: ndili6} from Pad\'{e} QCD.

\item From a completely different phenomenological perspective based on known masses of hadronic resonances, Greiner
and Schafer~\cite{Greiner94} had also argued that the transition from NPQCD to PQCD must occur rapidly within a small segment
of the energy scale they found to be  $1  \le  \sqrt{Q^2}  \le 3$ GeV such that NPQCD is definitely all of $Q \le 1$ GeV,
while all of $\sqrt{Q^2} \geq 3$ GeV is  PQCD.  Stated in our own terms, this Greiner-Schafer argument places the onset 
point of NPQCD/$D \chi SB$ in long distance QCD dynamics, in the range $1 \le \Lambda_{\chi}^2 \le 3$ GeV, or
$1.0 \le \Lambda_{\chi} \le 1.8$ GeV,  to be compared  with the values: $0.706 \le  \Lambda_{\chi}  \le 2.6$ GeV
for $N_{f} \le 6$,  we found from our Pad\'{e} QCD and  shown variously in Tables~\ref{tab: ndili5} and~\ref{tab: ndili6}.

\item On the specific issue of the confinement scale in QCD, Brown and Pennington~\cite{Brown88} obtained an
infrared plot based on considerations of the Schwinger-Dyson equation for the gluon propagator, showing that the
gluon propagator gets enormously enhanced and rises sharply in value to infinity in a Landau type cut-off behavior,
at a scale $Q \approx \Lambda_{\mathrm{QCD}}$ in the infrared QCD region.  This infrared  gluon propagator behavior
is normally  taken as indicative of the point of onset of confinement in QCD.  Its phenomenological scale found here  at 
$ Q = Q_{c} \approx \Lambda_{\mathrm{QCD}}$ is to be
compared with our $a_{4}$ or $Y_{2}$ plots shown variously in figs.~\ref{fig: ndili600a},~\ref{fig: ndili600b},
and figs.~\ref{fig: ndili605} to~\ref{fig: ndili607},  which rise sharly to infinity at exactly 
similar critical momentum points $Q_{c}(Y_{2}) \approx 296 MeV \approx \Lambda_{\mathrm{QCD}}$.

\item The whole question of the enhanced $Q^{-4}$ infrared singularity  behavior of the gluon propagator has actually 
become a central pivot for formulating  chiral symmetry breaking and confinement dynamics of QCD, and estimating the
the scale boundaries of these phenomena.  Gogohia and Magradze~\cite{Gogohia89}   used it to establish a close
interrelatedness and juxtaposed existence in QCD  of  dynamical  chiral symmetry  breaking and confinement, exactly
as  our Pad\'{e} QCD finds in  figs.~\ref{fig: ndili600} to~\ref{fig: ndili600b} and~\ref{fig: ndili605} 
to~\ref{fig: ndili609}.  Arbuzov~\cite{Arbuzov88}  working from the same premise of $Q^{-4}$ infrared singularity
behavior of the gluon propagator, derives various essential  attributes of spontaneous chiral symmetry breaking and other 
nonperturbative infrared QCD phenomena, and finds a momentum scale $k_{o}$ separating asymptotically free ultra violet
QCD from the  nonperturbative infrared domain  to be  $k_{o} = 700$ MeV for the specific flavor state $N_{f} = 0$ he 
considered.  This Arbuzov value compares again very well with our value $\Lambda_{\chi} = 706$ MeV for the same 
$N_{f} = 0$,  in the $R_{e+e-}(Q)$ values shown in Table~\ref{tab: ndili5}.  Other  recent 
studies~\cite{Gogohia99, Gogohia2000, Roberts94} of the nonperturbative QCD vacuum, assumed dominated by a $Q^{-4}$ gluon 
propagator  behavior, are found to  exhibit a two scale structure of chiral symmetry breaking and quark confinement
dynamics,  all strikingly similar to our Pad\'{e} infrared  QCD findings reported here. 

\item Finally Iwasaki and co-workers~\cite{Iwasaki92} found from their Lattice QCD studies that
dynamical chiral symmetry breaking and confinement in infrared QCD must be closely related or correlated, even when
they found no evidence that the two phenomena necessarily occur simultaneously. The implication for scale considerations,
is that the two phenomena are compatible with being  intrinsic two scale structure  or sequential 
phenomena of infrared QCD dynamics, exactly as we find  here with infrared Pad\'{e} QCD. 
What is more,  Iwasaki et. al. found that their two scales and phenomena of infrared  QCD, depend on the flavor number 
$N_{f}$ of the QCD system,  again exactly as our Tables~\ref{tab: ndili5} and~\ref{tab: ndili6}, and
our various Pad\'{e} QCD plots, explicitly show.

\end{enumerate}

Clearly, these varied physical QCD phenomenological results concerning infrared QCD dynamics and its critical
boundary  points, all consistently close to and in agreement with the Pad\'{e} QCD findings shown variously
in figs.~\ref{fig: ndili600} to~\ref{fig: ndili600b}, and figs.~\ref{fig: ndili605} to~\ref{fig: ndili600c}, as well as
Tables~\ref{tab: ndili5} and~\ref{tab: ndili6}, compel us to admit the Pad\'{e} infrared $(a_{3}, a_{4})$ structures 
as relating to physical  QCD.  In that case, we can make a number of important deductions from our work, concerning 
infrared dynamics of physical QCD as follows. 

\section{FAVORED FLAVOR STATES FOR QUARK CONFINEMENT IN QCD}
One immediate deduction or proposal we can make following from the above, is that confinement and chiral symmetry
breaking in QCD are caused by two distinct component  color forces or dynamics of QCD.
 Of our two  nonperturbative infrared couplant forces $Y_{2}$ and $Y_{1}$, the latter is responsible for and
identifiable with the onset and dynamics of chiral symmetry breaking in QCD, while the former $Y_{2}$ is  responsible for
quark confinement in QCD.  Following from the fact that the two component color forces are independent solutions 
$a_{3}$ and $a_{4}$ of our Pad\'{e} couplant equation, we assert as part of our finding, that the two nonperturbative 
infrared QCD phenomena, chiral symmetry breaking and quark confinement, are actually distinct 
physical phenomena.  They may however still be closely related in the sense that  one  phenomenon  provides a stimulus or
background that facilitates the occurrence of  the other phenomenon.

 The indication from our plots is that  dynamical chiral symmetry breaking
$(D \chi SB)$ occurs first at the higher infrared nonperturbative scale $Q_{c}(Y_{1})$, phenomenologically flooding the QCD
vacuum with $q \bar q$ pair condensates, which in turn provide a superconducting state of QCD vacuum, compatible with quark 
confinement in the manner advocated by a number of workers~\cite{Konishi2001,Giacomo98, Kondo2000}. However, what our work
now indicates is that  confinement itself does not actually occur (even after a $D \chi SB$) until the independent 
$Y_{2}$ confinement causing  color force has attained its own confinement scale at $Q = Q_{c}(Y_{2}) < Q_{c}(Y_{1})$.
In such a scenario, the possibility  arises that if these two scales are very far apart, the condensate background
effect caused by  $Y_{1}$ at scale  $Q_{c}(Y_{1})$ may become dissipated or ineffective  by the time we reach the
$Y_{2}$ confining scale at $Q_{c}(Y_{2})$.  Then confinement does not occur even though we are at scale $Q_{c}(Y_{2})$.

The details of this confinement mechanism  do not concern us here. What we show now is that our work
provides a strong evidence or indication that the above interrelatedness or correlation of confinement  with dynamical chiral 
symmetry breaking does exist in concrete terms in  QCD.  Here we return to  Table~\ref{tab: ndili5} and~\ref{tab: ndili6}
where we already computed the ratios $Q_{c}(Y_{2})/Q_{c}(Y_{1})$ as a measure of how close together or far apart the two 
nonperturbative infrared scales are. We now plot this $Q_{c}(Y_{2})/Q_{c}(Y_{1})$ correlation factor as a function of 
flavor number  and for different  spacelike and timelike observables.  The plots are shown  
in fig.~\ref{fig: ndili610}, and reveal some  interesting features.

We observe that in the spacelike cases the correlation ratio first rises with increasing flavor number from $N_{f} = 0$, and
reaches a peak at $N_{f} = 2$  or in the region $1 \le N_{f} \le 3$.  It then falls off rapidly for all $N_{f} \geq 4$ and
is practically zero near $N_{f} = 7, 8$.  It is well known in nature,  that the most stable hadrons and confined
quark states are the $(u,d)$ or $N_{f} = 2$ states of QCD.  These are the nucleons (protons and neutrons).  $N_{f} = 3$
confined quark states exist as higher baryon states that are only quasi stable.   There are no known stable or quasi stable 
confined quark states of flavor number $N_{f} \geq 4$.  Our fig.~\ref{fig: ndili610} appears therefore to be  providing us
a strikingly direct explanation and insight into the origin of this well known  state of affairs in nature, attributing it
to an interplay or some intrinsic interrelatedness  of dynamical chiral symmetry breaking and quark confinement mechanism 
in QCD.

In the case of the timelike $Q$ cases shown also in fig.~\ref{fig: ndili610},  the correlation factor
falls off progressively from $N_{f} = 0$ to $N_{f} = 8$, in a manner to suggest that the most readily confined
QCD states are the pure gluonic states or glue balls, rather than the  $1 \le N_{f}  \le 3$ states.  This would appear not 
to be so in nature but we are not able at the moment to reconcile this crucial  difference in the timelike and spacelike 
behaviors of fig.~\ref{fig: ndili610}, except that the difference is directly traceable to $\rho_{2}$ being positive or 
negative in the manner shown in Table~\ref{tab: ndili1}.  Apart from this difference in their $N_{f} \le 3$ behavior, the 
timelike and spacelike observables are fully in agreement from their correlation  plots shown in fig.~\ref{fig: ndili610}, that confinement becomes
rapidly less probable for $N_{f} \geq 4$, and that this decline arises from the intrinsic confinement scale $Q_{c}(Y_{2})$ 
becoming progressively less correlated with the intrinsic dynamical chiral symmetry breaking scale $Q_{c}(Y_{1})$ of QCD.

We remark that Iwasaki and co-workers~\cite{Iwasaki92} actually found in their Lattice QCD studies, that confinement and
dynamical chiral symmetry breaking cease to occur for $N_{f} \geq 7$, but occur  only for $N_{f} \le 6$.  Our work
and fig.~\ref{fig: ndili610} explain adequately what Iwasaki and co-workers observed in their Lattice QCD studies, and 
this close agreement between their work and ours, leads us once more to affirm that  our Pad\'{e} infrared  QCD results
have a physical reality, reproducing quite well infrared QCD dynamics from a  wide variety of phenomenological 
considerations, including  Lattice QCD simulations.

\section{FROZEN COUPLANT VERSUS INFRARED SLAVERY STRUCTURE OF CQCD}

 A second deduction for QCD we can make from our  Pad\'{e} plots  and features, is that while  
many of the phenomenological models we cited earlier in this paper, especially those dealing with ladder approximant 
Schwinger-Dyson equations, make mere guesses or assumptions about the form of the QCD coupling 
constant inside the CQCD regime,  $(\Lambda_{\mathrm{con}} \le Q \le \Lambda_{\chi})$,  some authors parameterizing 
it by a frozen coupling constant or a step function:

\begin{equation}\label {eq: ndili94}
g^2((q - p)^2)  = \theta (q^2 - p^2) g^2(q^2)  +  \theta (p^2 - q^2) g^2(p^2)      
\end{equation}

with \\
\begin{eqnarray}
\theta (q^2 - p^2)  &=& 1,   \mbox{if  $q^2 > p^2$}               \nonumber\\
                    &=& 0,  \mbox{if  $q^2 < p^2$}        \label{eq: ndili95}
 \end{eqnarray}
or in the form:
\begin{eqnarray}
g^2(q^2)   &=& \frac{2}{b \ln \left( \frac{(q_{c}^2 + q^2)}{\Lambda_{\mathrm{QCD}}^2} \right )}      \nonumber\\
           &=& \frac{2}{b \ln \left( \frac{ q^2}{\Lambda_{\mathrm{QCD}}^2} \right )},  for : q > q_{c}    \nonumber\\
           &=& \frac{2}{b \ln \left( \frac{q_{c}^2}{\Lambda_{\mathrm{QCD}}^2} \right )},  for : q \le q_{c}    \label{eq: ndili96}
 \end{eqnarray}
 
where  $q_{c}$ is some constant parameter identifiable with our $\Lambda_{\chi}$,
our Pad\'{e} model shows explicitly~\cite{ndili2001}  that  the $N_{f} \le 8$ flavor states are not infrared 
couplant freezing states of QCD.  Rather the two coupling constant solutions $a_{3}$ and $a_{4}$ that govern  QCD
dynamics in this CQCD infrared regime of QCD are together growing infinitely large  in the domain.  Such
infinitely growing coupling constant in the infrared region of QCD, is what is naturally needed to explain physical
confinement of colored quarks and gluons into hadrons. Therefore, our optimized Pad\'{e} QCD has this additional
positive side to it that it offers a more natural explanation of the origin and mechanism of quark confinement than 
some of the phenomenological models we cited in section 5.  Indeed the  $Q^{-4}$ enhanced infrared singularity structure of 
the gluon propagator, currently gaining ground~\cite{Gogohia99, Arbuzov88, Gogohia2000, Roberts94} as a necessary 
ingredient to reconcile QCD vacuum with  quark confinement, the linearly rising inter-quark potential law, and other
nonperturbative features of infrared QCD, has no place for an infrared fixed point in low flavor states of QCD.
The same conclusion that low flavored states of QCD do not manifest any infrared stable fixed point is found 
in several other places Refs.~\cite{Gardi98} to~\cite{Miransky97}, all rather contrary to
Refs.~\cite{Mattingly92, Mattingly94}.

\section{SUMMARY AND CONCLUSIONS}

Summarizing, we state that the two scale cut off momentum structure exhibited in the infrared region of QCD
by our optimized  Pad\'{e} couplant eqns.~(\ref{eq: ndili1}) and~(\ref{eq: ndili2}), is identifiable as marking the
points of onset of dynamical chiral symmetry breaking and  quark  confinement respectively, in infrared QCD. Taking
these two infrared nonperturbative phenomena of QCD as characteristically marking  the  beginning phase  and end phase
respectively, of the  CQCD regime of QCD, we can conclude, based on extensive correlations of our results with a wide range
of known phenomenological studies of low energy QCD, that our work actually constitutes a finding from within 
perturbative QCD approaches, of a previously unknown but long speculated infrared QCD window, marked at its two ends 
by a critical coupling constant and a critical momentum, and separating QCD dynamics in the cold (zero temperature vacuum state), 
into the three distinct phases of purely hadronic QCD at very low momentum transfers Q, a largely weak perturbative QCD 
phase at high Q, and finally an intermediate phase or regime  of strongly coupled yet quark-gluon dominated  QCD dynamics. 

Traditionally in perturbative QCD (PQCD), one believes that an impenetrable wall at $Q = \Lambda_{\mathrm{QCD}}$  shuts
out nonperturbative QCD and hadron physics from any view or access from the PQCD end. This apart, it is held that
much of PQCD remains accessible in the domain stretch from the far ultraviolet (UV) QCD, until close to the Landau 
cut off wall at $Q = \Lambda_{\mathrm{QCD}}$.  What our work shows is that while the Landau wall at
$Q = Q_{c}(Y_{2}) \approx  296 MeV \approx  \Lambda_{\mathrm{QCD}}$ remains impenetrable  from the PQCD end, much of the most
important aspects of nonperturbative QCD dynamics including spontaneous chiral symmetry breaking and the  confinement
of quarks into hadrons, actually lies in a buffer zone (CQCD) that  leads up from the far UV QCD to the Landau wall.
Beyond the Landau wall lies pure hadron physics, and our work provides explicit boundary values both in critical momentum
and critical QCD coupling constant, separating these three  phases of QCD, as delineated from the PQCD  beta function
parameters and renormalization group equations of a running QCD coupling constant. 

We believe the same CQCD buffer zone just delineated,  is what is  currently being  observed as Quark Gluon Plasma (QGP) from 
the hadron end at high temperature (non-vacuum QCD state), in heavy ion experiments at CERN and Brookhaven (RHIC). As
such, it becomes immediately possible  to ask how  the domain boundaries and dynamical characteristics of this 
intermediate CQCD/QGP phase of QCD get modified or shifted  at two widely separated temperatures and energy density, 
represented by the heavy ion experiments on the one hand,  and our zero temperature vacuum QCD studies on the other hand.
This comparison will await further developments and evolution in the QGP plasma experiments and the  various Lattice
and phenomenological computations relating thereto.

\begin{figure}
\scalebox{1.0}{\includegraphics{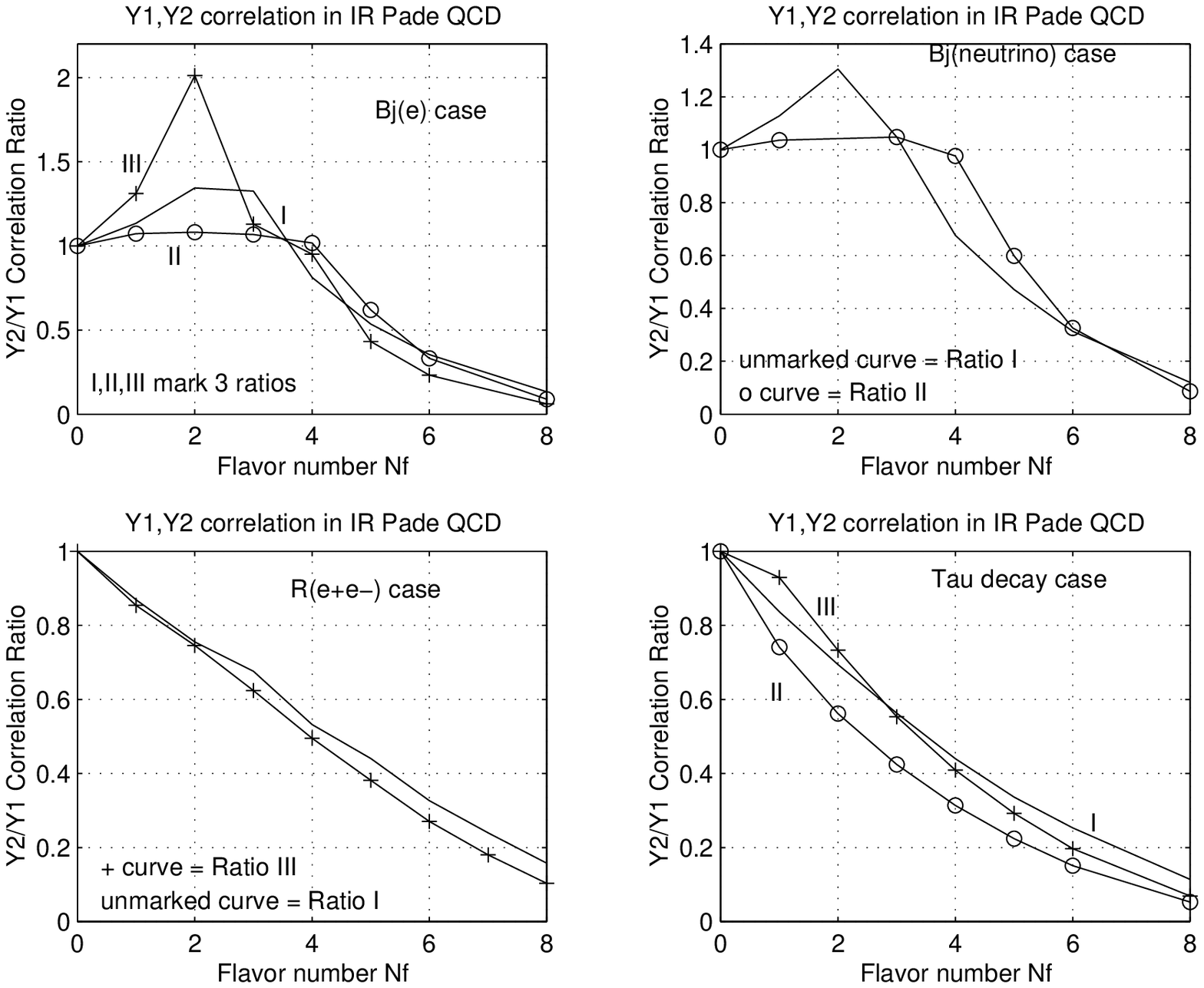}}
\caption{ A plot showing the strong correlation between dynamical chiral symmetry breaking and quark confinement 
in QCD as revealed in Pad\'{e} infrared couplant structures.  Quark confinement is seen to be most favored in
$N_{f} \le 3$ states  with a peak at $N_{f} = 2$, but drops rapidly to zero for $N_{f} \geq 4$, exactly as one
finds in nature. Each curve was normalized to its $N_{f} = 0$ flavor state.}
\label{fig: ndili610}
\centering
\end{figure}

Focusing on our own boundary  values of the CQCD regime computed here at zero temperature, we found that these values 
in general vary with flavor as seen from Tables~\ref{tab: ndili5} and~\ref{tab: ndili6}, and from our various figures and 
plots. We have  interpreted this $Q_{c}(Y_{2})/Q_{c}(Y_{1)}$ flavor behavior to be a maifestation of a  phenomenologically
known fact that dynamical chiral symmetry breaking and quark confinement are closely related infrared QCD phenomena.  Our
specific finding is that the more widely separated  their set  scales are, the less likely the corresponding flavor state 
of QCD will be a quark confinement state. We actually found evidence from our fig.~\ref{fig: ndili610} that quark 
confinement occurs mainly in the low flavor states $N_{f} \le 3$, with a peak at $N_{f} = 2$, but becomes  progressively 
non-existent or improbable in the higher flavor states $N_{f} = 4, 5, 6, 7, 8$.  It turns out that this is actually 
what one finds in nature.

After we had completed this work, our attention was drawn to a preprint by Elias and co-workers~\cite{Elias2001} in which
they used their own method of  denominator versus numerator zeros of Pad\'{e} QCD beta function, to estimate the location
and scale value of the point where perturbative QCD cuts off from the infrared region. They found only one scale cut-off point 
at a critical momentum $\mu_{c} \approx 1.0$ GeV.  This is  in contrast to our two scale cut off structure presented variously in 
this paper,  and anticipated in the several phenomenological models we cited in section 5 of this paper. We make two comments
in relation to this work of Elias et. al.  The first is that the method of denominator versus numerator zero of Pad\'{e}
beta function appears not adequate to reveal the multicomponent couplant solutions of Pad\'{e} QCD and consequently the 
two scale infrared structure.   This fact  was already highlighted in our earlier paper~\cite{ndili2001}.  One needs to work with
effective charges and renormalization group invariants or optimized Pad\'{e} QCD.  The second comment we make is one of
noting that the one scale cut off found by Elias et. al. at $\mu_{c} \approx 1.0$ GeV, is comparable to our chiral symmetry
breaking scale  $\Lambda_{\chi}  \approx 1.44$ GeV.  Since Elias et. al. worked with the higher Pad\'{e} approximants
$[2|1], [1|2], [3|1], [1|3], [2|2]$, as against our $[1|1]$ Pad\'{e} approximant, the close agreement between our 
onset scale and theirs, provides a good indication that our results and features presented here will  hold irrespective of the
particular Pad\'{e} approximant used, exactly as we already inferred in paper~\cite{ndili2001}.  One could
 of course repeat directly the various computations reported in this paper, using this time,  these higher Pad\'{e} 
approximants to verify that the results remain robust.   The draw back is that the higher renormalization scheme invariants
$\rho_{3}, \rho_{4}...$ needed for such computations, are not presently known or computed 
perturbatively for the various QCD physical observables reported in this paper.\\

{\bf ACKNOWLEDGEMENTS} \\
The author thanks Dr. Billy Bonner, Director T.W. Bonner Nuclear Laboratory, Rice University, Houston, for 
hospitality. He is particularly  grateful to Dr. Awele Ndili of Stanford University, California,
for much technical advice on computing and software.  He  thanks the following colleagues
who kindly read an  earlier version of this paper and made  suggestions for improvement: Dr. V. Miransky
and Dr. V. Elias  of Western Ontario Canada,  Dr. B. McKellar of Melbourne Australia,
Dr. K. Higashijima of Osaka Japan, and Dr. V. Gogokhia of Budapest Hungary. Finally, the author thanks Fredna Associates 
for financial support that made this work possible.

\newpage

\begin{table}
\caption{Computed $\rho_{2}(N_{f})$ values for various spacelike and timelike QCD observables}\label{tab: ndili1}

\begin{tabular}{|l|l|l|l|l|l|l|}\hline
Flavor & $\rho_{2}^{Bj(e)}$ & $\rho_{2}^{Bj(\nu)}$  &  $\rho_{2}^{GLS(\nu)}$  &  $\rho_{2}^{(e^+e^-)}$  & $\rho_{2}^{(\tau)}$  & $\rho_{2}^{(H)}$\\
Number  &          &                          &                         &                         &                      &   \\
$N_{f}$  &          &                          &                         &                         &               &  \\
\hline

0   &  17.9244   &  15.9567  &  17.9244  &  -8.410032589173554   &  1.8089  &  -17.8458  \\

1  &  13.7474  &  12.3442   &  13.3343   & -9.996607149709793   &  -0.5654   & -18.9820   \\

2  &  9.6051  &  8.7025    &  8.7787   & -10.91120013471925     &  -2.9879    &  -20.1684   \\

3  &  5.4757  &  5.0067    &  4.2361  & -12.20710268197531     &  -5.4756     &   -21.4108  \\

4  &  1.3304  &   1.2238  &  -0.3223  & -13.90995802777778    &   -8.0508    &  -22.7170  \\

5  & -2.8695  &  -2.6906   &  -4.9354  &  -15.49181836878120   &   -10.7438   & -24.0975  \\

6  & -7.1775  &  -6.7981   &  -9.6566  &  -17.66469557734694   &  -13.5963   &  -25.5668  \\

7  & -11.6698   &  -11.1857  &  -14.5620  &  -19.78668878025239    &  -16.6675   &  - 27.1456   \\

8  &  -16.4581  &  -15.9819  &  -19.7635   & -22.74511792421761     &  -20.0446  &  -28.8640   \\

9   & -21.7139   &  -21.3833  & -25.4325  &  -25.96983971428571     &  -23.8613   &  -30.7687 \\

10  & -27.7143   &  -27.7079  & -31.8461  &  -30.64825148592373     &  -28.3334  &  -32.9345  \\

11  & -34.9379   &  -35.5045  &  -39.4829  & -36.70527878387512    & -33.8340   &  -35.4912   \\

12  & -44.2885   &  -45.8101  &  -49.2467  &   -46.58505774333333   &  -41.0672  & - 38.6836   \\

13  & -57.7031   &  -60.8467  & -63.0745   &  -63.56012999671129    &  -51.5427  &  -43.0365  \\

14  &  -80.2178  &  -86.3914   &  -86.0023  & -101.9145678055556    &  -69.1847  &  -49.9139 \\

15  &  -130.2979  &  -143.6271  &  -136.4956  &  -229.8874551066667     & -108.3829  & -64.3179  \\

16  & -374.1328  &  -423.2004   & -380.7437   &  -1724.404563921111     &  -298.6440  & -131.2671    \\
\hline
\end{tabular}
\end{table}

\begin{table}
\caption{$Y_{1}$ structural parameters $Q_{\mathrm{min}},  a_{1}(Q_{\mathrm{min}}),  Q_{\mathrm{max}}$ and
$a_{3}(Q_{\mathrm{max}})$ for Bjorken sum rule $\rho_{2}^{Bj(e)}$ values shown in Table~\ref{tab: ndili1}
in optimized $[1|1]$ Pad\'{e} QCD. The $N_{f} = 0, 1, 2$ cases were computed using $\sqrt{|D|}.$}\label{tab: ndili2}
\begin{center}

\begin{tabular}{|c|l|l|l|l|}\hline
Flavor Number $N_{f}$ & $Q_{\mathrm{min}}$ (GeV) & $a_{1}(Q_{\mathrm{min}})$ & $Q_{\mathrm{max}}$ (GeV) & $a_{3}(Q_{\mathrm{max}})$\\
\hline

 0   &  16.3391   &  0.0812   &  633.07   &  0.10006\\

 1   &  12.5361  &   0.0936   &   444.10   &  0.1174\\

 2   &  8.0622   &  0.1164   &  325.33   &  0.15139\\

 3   &  6.11284  &   0.13944   &   364.81   &  0.18538\\

 4   &  11.1934   &  0.12027  &   238.4  &  0.1477\\

 5   &  17.012   &  0.1149  &  204.84  &   0.1344\\

 6   &  24.2354   &   0.11425   &  161.30   &  0.1278\\

 7   &  32.592   &  0.11665   &  115.8   &  0.1243\\

 8   &  36.214   &  0.12191   &  37.56  &   0.12235\\

 9   &  6.910   &   0.12087   &  36.279   &  0.1303\\

 10   &  0.3015   &   0.11868  &  62.46  &   0.1428\\

 11   &  1.036$(10^{-3})$   &   0.1143   &  250.952   &  0.16116\\

 12   &  8.2160$(10^{-7})$   &   0.1059  &   7.7850$(10^{+3})$   &  0.1900\\

 13   &  9.029$(10^{-14})$   &   0.0918  &   7.8324$(10^{+7})$    &  0.2395\\

 14   &  9.752$(10^{-35})$  &   0.07130  &   7.8451$(10^{+20})$  &   0.3652\\

 15   &  1.281$(10^{-101})$  &   0.0448   &  $\approx 1.0(10^{+85})$   &  $> 0.80$\\

 16   &  2.250$(10^{-176})$    &  0.01520   &  $\gg 10^{+308}$   &   $\gg  1.0$\\
\hline
\end{tabular}
\end{center}
\end{table}

\begin{table}
\caption{Bjorken neutrino sum rule $Y_{1}$ structural parameters $Q_{\mathrm{min}},  a_{1}(Q_{\mathrm{min}}),
Q_{\mathrm{max}}$ and $a_{3}(Q_{\mathrm{max}})$ computed using the $\rho_{2}^{Bj(\nu)}$ values shown in 
Table~\ref{tab: ndili1}, in optimized $[1|1]$ Pad\'{e} QCD. The $N_{f} = 0, 1, 2$ cases were computed using 
$\sqrt{|D|}$.}\label{tab: ndili3}
\begin{center}

\begin{tabular}{|c|l|l|l|l|}\hline
Flavor Number $N_{f}$   &  $Q_{\mathrm{min}}$ (GeV)  &  $a_{1}(Q_{\mathrm{min}})$  &    $Q_{\mathrm{max}}$  (GeV)  &  $a_{3}(Q_{\mathrm{max}})$\\
\hline

 0   &  16.271   &  0.0868   &  404.4   &  0.1087\\

 1   &  12.432  &   0.1005   &   403.6   &  0.1284\\

 2   &  7.63   &  0.1285   &  326.6   &  0.1724\\

 3   &  8.5585  &   0.1312   &   266.8   &  0.171\\

 4   &  13.5895  &  0.1194  &   732.4  &  0.1464\\

 5   &  19.388  &  0.1159  &  281.4  &   0.1357\\

 6   &  26.4586   &   0.1159   &  343.3    &  0.1299\\

 7   &  34.4886   &  0.1186   &  113.8   &  0.1265\\

 8   &  37.52   &  0.1237   &  39.72  &   0.12415\\

 9   &  5.117   &   0.1219   &  37.18   &  0.1315\\

 10   &  0.2723   &   0.1187  &  64.03  &   0.1430\\

 11   &  1.678$(10^{-3})$   &   0.1133   &  260.28   &  0.1590\\

 12   &  8.130$(10^{-8})$   &   0.10402  &   8.2617$(10^{+3})$   &  0.1835\\

 13   &  9.620$(10^{-15})$   &   0.0897  &   8.592$(10^{+7})$    &  0.2255\\

 14   &  6.69$(10^{-42})$  &   0.06957  &   8.9984$(10^{+20})$  &   0.3253\\

 15   &  3.440$(10^{-83})$  &   0.0441   &  1.02687$(10^{+85})$   &  1.465\\

 16   &  4.90$(10^{-307})$    &  0.015040   &  $\gg 10^{+305}$   &  $\gg 10.0$\\
\hline
\end{tabular}
\end{center}
\end{table}

\begin{table}
\caption{Tau($\tau$) Hadronic decay $Y_{1}$ structural parameters $Q_{\mathrm{min}},  a_{1}(Q_{\mathrm{min}}), 
Q_{\mathrm{max}}$ and $a_{3}(Q_{\mathrm{max}})$ computed using the $\rho_{2}^{\tau}$ values shown in 
Table~\ref{tab: ndili1}, in optimized $[1|1]$ Pad\'{e} QCD. All $D$ values in this case, including $N_{f} = 0$ state,
are positive.}\label{tab: ndili4}
\begin{center}

\begin{tabular}{|c|l|l|l|l|}\hline
Flavor Number $N_{f}$   &  $Q_{\mathrm{min}}$ (GeV)  &  $a_{1}(Q_{\mathrm{min}})$  &    $Q_{\mathrm{max}}$  (GeV)  &  $a_{3}(Q_{\mathrm{max}})$\\
\hline

 0   &  26.6430   &  0.0773   &  765.30   &  0.0942\\

 1   &  32.0952  &   0.07825   &   1697.95   &  0.0942\\

 2   &  38.7402   &  0.07977   &  749.06   &  0.09475\\

 3   &  46.851  &   0.08194   &   785.50   &  0.0959\\

 4   &  56.716  &  0.0849  &   813.4  &  0.09772\\

 5   &  68.5620  &  0.08893  &  660.9  &   0.10015\\

 6   &  82.1620   &   0.09422   &  547.18   &  0.1033\\

 7   &  95.261   &  0.10123   &  267.0   &  0.10692\\

 8   &  92.2863   &  0.11058   &  95.54  &   0.11094\\

 9   &  15.2460   &   0.1147   &  81.01   &  0.1232\\

 10   &  0.591   &   0.1172  &  122.64  &   0.1407\\

 11   &  3.5330$(10^{-3})$   &   0.1164   &  435.71   &  0.1654\\

 12   &  1.2110$(10^{-6})$   &   0.11016  &   1.2066$(10^{+4})$   &  0.2035\\

 13   &  2.015$(10^{-14})$   &   0.0964  &   1.09640$(10^{+8})$    &  0.2728\\

 14   &  3.505$(10^{-33})$  &   0.0745  &   1.00163$(10^{+21})$  &   0.4690\\

 15   &  2.636$(10^{-105})$  &   0.0463   &  $\geq 1.0(10^{+88})$   &  $\gg 1.0$\\

 16   &  $1.06 10^{-295}$    &  0.0153   &  $\gg 10^{+305}$   &   $\gg 1.0$\\
\hline
\end{tabular}
\end{center}
\end{table}

\begin{table}
\caption{CQCD boundary  momenta $Q_{c}(Y_{1})$ and $Q_{c}(Y_{2})$  computed for various physical observables
or effective charges in the $[1|1]$ optimized Pad\'{e} QCD. The $Q_{c}(Y_{1}), Q_{c}(Y_{2})$ values are taken from 
figs.~\ref{fig: ndili600a},~\ref{fig: ndili600b}, and~figs.~\ref{fig: ndili605} to~\ref{fig: ndili609}, at the
three strategic couplant points discussed in the paper. Ratio I = $Q_{on}(Y_{2})/Q_{0n}(Y_{1})$, while Ratio III
= $Q_{off}(Y_{2})/Q_{off}(Y_{1})$. Ratio II was evaluated at the IR attractor point of the
 $Y_{2}$ curve where this exists.}\label{tab: ndili5}
\begin{center}

\begin{tabular}{|l|l|l|l|l|l|l|l|}\hline
QCD   &  Flavor    &  $Q_{on}(Y_{2})$   &  $Q_{on}(Y_{1})$  &  Ratio & $Q_{off}(Y_{2})$ &  $Q_{off}(Y_{1})$  & Ratio \\
observable & $N_{f}$  &   GeV            &      GeV            &  I    &  GeV               &  GeV               &  III/II \\
\hline
 Bj(e)  &  0   &  1.0   &  2.375  &  0.4211  &  0.8791  &  0.9465  &  0.9287 \\
       
        &  1   &  1.195  &  2.50  &  0.4780  &  1.0956  &  0.900   &  1.2173 \\ 
        
        &  2   &  1.60  &  2.825  &  0.5664  &  1.5661   &  0.8380  &  1.8689 \\       
       `
        &  3   &  1.815  &  3.25  &   0.5585  &  0.85   &  0.81   &  1.0494 \\

        &  4   &  1.09   &  3.186  &  0.3421   &  0.78278  &  0.88625   &  0.8833 \\

        &  5   &  0.8125  &  3.60  &  0.2257   &  0.3953   &  0.98629   &  0.4008 \\

        &  6   &  0.6340  &  4.25  &  0.1492  &  0.2450   &  1.1392   &  0.2150 \\

        &  8   &  0.440  &  7.80  &  0.0564  &  0.1084   &   1.900  &  0.0571  \\
\hline
 Bj($\nu$) &  0   &  1.4474  &  2.9460  &  0.4913  &  1.349312  &  1.9650  &  0.6867 \\
 
           &  1   &  1.7177  &  3.10    &  0.5541  &  1.66233  &  2.3360  &  0.7116  \\
           
           &  2   &  2.4460  &  3.8150    &  0.6412  &  -  &  -  &  -  \\
              
           &  3   &  1.9365  &  3.760  &   0.5150  &  1.856   &  2.5780   &  0.7199 \\

           &  4   &  1.30  &  3.910  &  0.3325  &  0.911941  &  1.3602   &  0.6704 \\

           &  5   &  0.95  &  4.10  &  0.2317  &  0.4732   &  1.15174   &  0.4109 \\

           &  6   &  0.75  &  4.870  &  0.1540  &  0.290   &  1.2941   &  0.2241 \\
           
           &  8   &  0.4710  &  8.0    &  0.0589  &  0.1225  &  2.0593  &  0.0595  \\   
\hline
$R_{e+e-}(Q)$  &  0   & 0.3700  &  1.700  &  0.2176  &  0.2364  &  0.7064  &   0.3347 \\

               &  1   & 0.350  &  1.850  &  0.1892  &  0.20996  &  0.7345  &  0.2859 \\

               &  2  &  0.337  &  2.05  &  0.1644  &  0.19005   &  0.7611   &  0.2497 \\

               &  3   &  0.320  &  2.175  &  0.1471  &  0.16717  &  0.80043   &  0.2089 \\

               &  4   &  0.3037  &  2.625  &  0.1157  &  0.1425  &  0.8599  &   0.1657 \\
        
               &  5   &  0.2875  &  3.00  &  0.0958  &  0.1204  &  0.9420   &  0.1278 \\

               &  6  &  0.270  &  3.800  &  0.0711  &   0.09771   &  1.07991   &  0.0905 \\
               
               &  7  &  0.250  &  4.80  &  0.0521  &  0.079425  &  1.320  &  0.0602  \\
               
               &  8  &  0.249  &  7.25  &  0.0343  &  0.06250  &  1.8115  &  0.0345  \\
\hline
$\tau$ decay   &  0  &  1.2446  &  3.373  &  0.369  &  1.07444  &  1.340  &  0.8018  \\

$R^{(\tau)}$   &  1  &  1.1341  &  3.674  &  0.3087  &  0.8308  &  1.3980  &  0.5943  \\
               
               &  2   &  1.025  &  4.0  &  0.2562  &  0.6615   &  1.468   &  0.4506 \\

               &  3   &  0.9340  &  4.50  &  0.2076  &  0.5305  &  1.560   &  0.3401 \\

               &  4   &  0.850  &  5.237  &  0.1623  &  0.42325   &  1.6825   &  0.2516 \\
               
               &  5   &  0.7756  &  6.250  &  0.1241  &  0.3331  &  1.8540  &  0.1796 \\

               &  6   &  0.700  &  7.50  &  0.0933  &  0.260   &  2.146   &  0.1212 \\
               
               &  8   &  0.6170  &  14.70  &  0.0420  &  0.15226  &  3.590  &  0.0424  \\
\hline        
\end{tabular}
\end{center}
\end{table}

\begin{table}
\caption{Further data on CQCD boundary  momenta $Q_{c}(Y_{1})$ and $Q_{c}(Y_{2})$  computed for various physical observables
or effective charges in the $[1|1]$ optimized Pad\'{e} QCD. The $Q_{c}(Y_{1}), Q_{c}(Y_{2})$ values are taken from 
figs.~\ref{fig: ndili600a},~\ref{fig: ndili600b}, and~figs.~\ref{fig: ndili605} to~\ref{fig: ndili609}, at the
three strategic couplant points discussed in the paper. Ratio I = $Q_{on}(Y_{2})/Q_{0n}(Y_{1})$, while Ratio III
= $Q_{off}(Y_{2})/Q_{off}(Y_{1})$. Ratio II was evaluated at the IR attractor point of the
 $Y_{2}$ curve where this exists.}\label{tab: ndili6}
\begin{center}

\begin{tabular}{|l|l|l|l|l|l|l|l|}\hline
QCD   &  Flavor    &  $Q_{on}(Y_{2})$   &  $Q_{on}(Y_{1})$  &  Ratio & $Q_{off}(Y_{2})$ &  $Q_{off}(Y_{1})$  & Ratio \\
observable & $N_{f}$  &   GeV            &      GeV            &  I    &  GeV               &  GeV               &  III/II \\
\hline
 Bj(e)  &  0   &     &    &    &  0.879030  &  1.36  &  0.6463 \\
       
        &  1   &    &    &    &  1.0956  &  1.58    &  0.6934 \\ 
        
        &  2   &    &    &    &  1.5661   &  2.24  &  0.6992 \\       
       `
        &  3   &    &    &    &  1.7941495   &  2.600   &  0.6906 \\

        &  4   &    &    &    &  0.782782  &  1.190   &  0.6578 \\

        &  5   &    &    &    &  0.3953   &  0.98629   &  0.4008 \\

        &  6   &    &    &   &  0.2450   &  1.1392   &  0.2150 \\

        &  8   &    &    &    &  0.1084   &   1.900  &  0.0571  \\
\hline
$\tau$ decay   &  0    &    &    &    &  1.021127  &  1.6610  &  0.6148  \\

$R^{(\tau)}$   &  1    &   &    &    &  0.829744217 &  1.45260  &  0.5712  \\
               
               &  2   &    &    &    &  0.6615   &  1.468   &  0.4506 \\

               &  3   &    &    &    &  0.5305  &  1.560   &  0.3401 \\

               &  4   &    &    &    &  0.42325   &  1.6825   &  0.2516 \\
               
               &  5   &   &    &    &  0.3331  &  1.8540  &  0.1796 \\

               &  6   &    &    &    &  0.260   &  2.146   &  0.1212 \\
               
               &  8   &    &    &    &  0.15226  &  3.590  &  0.0424  \\
\hline        
Higgs case     &  0    &  0.6212  &  4.13  &  0.1504    &  0.37228  &  1.7615  &  0.2113  \\

$R^{(H)}$      &  2   &  0.600   &  5.50   & 0.1091   &  0.3180   &  2.0644   &  0.1540 \\

               &  3   &  0.600   &  6.50  & 0.0923   &  0.28940  &  2.2825   &  0.1268 \\

               &  4   & 0.550   & 7.10   &  0.0775  &  0.26075   &  2.5868   &  0.1008 \\
               
               &  5   &  0.550  &  9.00  & 0.0611    &  0.23280  &  3.0260  &  0.0769 \\

               &  6   & 0.570   &  12.50  &  0.0456   &  0.2065   &  3.720   &  0.0555 \\
\hline        
\end{tabular}
\end{center}
\end{table}

\newpage
\listoftables

\newpage
\listoffigures


\newpage

\begin{thebibliography}{99}

\bibitem{Leutwyler94}  H. Leutwyler, Ann.  Phys. (N.Y.), 235, 165 (1994); also  J. Gasser and H. Leutwyler,
                      Ann. Phys. (N.Y.) 158, 142 (1984), and  Nucl. Phys. B250,  465 (1985)


\bibitem{Weinberg79}  S. Weinberg, Physica, A96, 327  (1979).


\bibitem{Nambu61}  Y. Nambu and G. Jona-Lasinio, Phys. Rev. 122, 345 (1961)


\bibitem{Miransky93}  V.  A.  Miransky,  \emph{Dynamical  Symmetry  Breaking in Quantum Field Theories}, Chapter 12,
                       (World Scientific Publishers, Singapore, 1993).


\bibitem{Higashijima91}  K.  Higashijima,   Suppl.  Prog.  Theor.  Phys.  104, 1  (1991) ;  also   K. Higashijima, in
                   \emph{Proceedings of the KEK Summer Institute on High Energy Phenomenology}, KEK 91-8, August (1990), and
                   Phys. Rev. D29, 1228 (1984).


\bibitem{Manohar84}  A. Manohar and H. Georgi, Nucl. Phys.  B234,  189  (1984).


\bibitem{Shuryak81}  E. Shuryak, Phys. Lett. 107B, 103 (1981)


\bibitem{Baal98}  P. van Baal (Editor), \emph{Confinement, Duality, and Nonperturbative Aspects of QCD}, 
       Proceedings NATO Advanced Science Institutes Series, Physics, Vol. 368, Plenum Press, New York, (1998)


\bibitem{Bhattacharya2001} T. Bhattacharya, R. Gupta and A. Patel (Editors),  \emph{Lattice 2000, Proceedings of the 
XVIIIth Intern Symposium on Lattice Field theory, Bangalore, India}, Nucl. Phys. B (Proc. Suppl), Vol.94, North
 Holland  Publishers, 2001.



\bibitem{ndili2001}  F. N. Ndili, Phys. Rev. D64, 014018 (2001). 

 
\bibitem{Gross73}  D.  J.  Gross and F.  Wilczek, Phys.  Rev.  Lett.  30, 1343 (1973);\\ 
 H.  D.  Politzer, Phys.  Rev.  Lett.  30, 1346  (1973).


\bibitem{Caswell74}  W.  E.  Caswell, Phys.  Rev.  Lett.  33, 244  (1974);\\  D.  R.  T.  Jones, Nucl.  Phys.  B75, 531 (1974);\\
    E.  S.  Egorian and O.  V.  Tarasov, Teor.  Mat.  Fiz.  41, 26 (1979)


\bibitem{Tarasov80} O.  V.  Tarasov, A.  A.  Vladimirov and A.  Yu.  Zharkov, Phys.  Lett.  B93, 429  (1980);\\  
    S.  A.  Larin and J.  A.  M.  Vermaseren, Phys.  Let.  B303, 334 (1993).


\bibitem{Ritbergen97}  T.  van  Ritbergen, J.  A.  M.  Vermaseren  and S.  A.  Larin, Phys. Lett. B400, 379  (1997); 
    J.  A.  M.  Vermaseren, S.  A.  Larin and T.  van Ritbergen,  Phys.  Lett.  B405, 327  (1997).


\bibitem{Stevenson81}  P. M. Stevenson, Phys.  Rev.  D23, 2916 (1981).


\bibitem{Stevenson86}  P.  M.  Stevenson, Phys.  Rev.  D33, 3130 (1986) 


\bibitem{Dhar83}   A.  Dhar, Phys.  Lett.  128B, 407  (1983).


\bibitem{Dhar84}   A.  Dhar and V.  Gupta, Phys.  Rev.  D29, 2822  (1984).


\bibitem{Grunberg84}  G. Grunberg, Phys.  Rev.  D29, 2315 (1984); and Phys. Lett. 95B, 70 (1980).


\bibitem{Kazakov85}  D. I. Kazakov and D. V. Shirkov, Sov. J. Nucl. Phys. 42, 487 (1985).


\bibitem{Chyla89}  J. Chyla,  Phys.  Rev.  D40, 676 (1989).


\bibitem{Bjorken66}  J. D. Bjorken, Phys.  Rev.  148,  1467 (1966);  Phys. Rev. D1, 1376 (1970).


\bibitem{Larin91a}  S. A. Larin and J. A. M. Vermaseren, Phys. Letts. B 259,  345 (1991).


\bibitem{Larin91b}  S. A. Larin, F. V. Tkachov and  J. A. M. Vermaseren, Phys. Rev. Letts. 66,  862  (1991).


\bibitem{Gross69}  D. J. Gross and C. H. Llewellyn Smith, Nucl. Phys. B 14,  337 (1969).


\bibitem{Chyla92}  J. Chyla, and A. L. Kataev  Phys. Letts. B 297,  385 (1992).


\bibitem{Gorishny91a} S. G. Gorishny, A.L. Kataev and  S. A. Larin, Phys. Letts. B 259,  144 (1991).


\bibitem{Surguladze91}  L. R. Surguladze  and  M. A. Samuel, Phys. Rev. Letts. 66,   560 (1991).


\bibitem{Chetyrkin79} G. Chetyrkin,  A.L. Kataev and  F. V. Tkachov, Phys. Letts. B85,  279 (1979).


\bibitem{Dine79}  M. Dine and J. Sapirstein. Phys. Rev. Lett. 43,  668 (1979).


\bibitem{Celmaster80} W. Celmaster and R. Gonsalves, Phys. Rev. Lett.  44,  560  (1980).


\bibitem{Chyla91}  J. Chyla, A. Kataev and  S. A. Larin, Phys. Letts. B267,  269 (1991).


\bibitem{Samuel91} M. Samuel and L. R. Surguladze, Phys. Rev. D44, 1602 (1991).


\bibitem{Braaten92}  E. Braaten, S. Narison and A pich, Nucl. Phys. B 373, 581 (1992).


\bibitem{Gorishny91b} S. G. Gorishny, A.L. Kataev,  S. A. Larin and L. R. Surguladze, Phys. Rev. D43,  1633 (1991).


\bibitem{Chetyrkin97} K. G. Chetyrkin, Phys. Letts. B390,  309 (1997).


\bibitem{Fomin83}  P. I. Fomin, V. P. Gusynin,  V. A. Miransky and Yu. A. Sitenko, Rivista Nuovo Cimento, 6, 1 (1983); 
                   also, P. I. Fomin and V. A. Miransky, Particles \& Nuclei, 16, 469 (1985);  
                   V. P. Gusynin and V. A. Miransky, Phys. Lett. B191, 141 (1987).

\bibitem{Atkinson88}  D. Atkinson,  P. W. Johnson and K. Stam, Phys. Lett. B201,  105, (1988); also
                     Phys. Rev. D37, 2996 (1988).


\bibitem{Roberts90} C. D. Roberts and B. H. J. McKeller,  Phys. Rev. D41, 672 (1990).


\bibitem{Atkinson87}  D. Atkinson and P. W. Johnson, J. Math. Phys. 28, 2488 (1987); also Phys. Rev. D37, 2290 (1988);
                      Phys. Rev. D37, 2296 (1988), and Phys. Rev. D35, 1943 (1987).                   


\bibitem{Guo97}  X.  H. Guo and T. Huang,  Nuovo Cimento 110A, 799 (1997); and 
                Intern. J. Mod. Phys. A9, 499 (1994)


\bibitem{Kogut82}  J. Kogut, M. Stone, H.W. Wyld, J. Shigemitsu, S. H. Shenker, and D.K. Sinclair, Phys. Rev. Lett.
                 48, 1140 (1982) 
                

\bibitem{Kogut83}  J. Kogut, M. Stone, H.W. Wyld, W. R. Wild, J. Shigemitsu, S. H. Shenker, and D.K. Sinclair,
 Phys. Rev. Lett. 50, 393 (1983) 
                

\bibitem{Gogohia90}  V. Sh. Gogohia, Gy. Kluge and B. A. Magradze, Phys. Lett. B244, 68 (1990)


\bibitem{Gogohia94}  V. Sh. Gogohia, Intern. J. Mod. Phys. A9, 759  (1994); also Phys. Lett. B224, 177 (1989), and
                     Phys. Rev. D40, 4157  (1989).

\bibitem{Gogohia99}  V. Sh. Gogohia, Gy. Kluge, H. Toki and T. Sakai, Phys. Lett. B453, 281 (1999); also
                    Intern. J. Mod. Phys. A15, 45 (2000)

\bibitem{Greiner94}  W. Greiner and A Schafer, \emph{Quantum Chromodynamics}, (Springer Verlag, Berlin - Heidelberg, Chaps 7 \& 8, 1994).


\bibitem{Brown88}  N. Brown and M. R. Pennington, Phys. Lett. B202, 257 (1988); Erratum B205, 596(E) (1988). Also
                   Phys. Rev. D38,  2266  (1988), and Phys. Rev. D39, 2723 (1989).

\bibitem{Gogohia89}  V. Sh. Gogokhia and B. A. Magradze, Phys. Lett. B217, 162 (1989) 


\bibitem{Arbuzov88}  B. A. Arbuzov, Sov. J. Part. \& Nuclei 19, 1  (1988)


\bibitem{Gogohia2000}  V. Sh. Gogohia,  Phys. Lett. B485, 162 (2000) ; and hep-ph/0102261.


\bibitem{Roberts94}  C. D. Roberts and A. G. William, Prog. Part. \& Nucl. Phys. 33.  477 (1994). 


\bibitem{Iwasaki92}  Y. Iwasaki, K. Kanaya, S. Sakai and T. Yoshie, Phys. Rev. Lett. 69, 21 (1992)


\bibitem{Konishi2001}  K. Konishi and  K. Takenaga, Phys. Lett. B508, 392 (2001). 


\bibitem{Giacomo98}  A. Di Giacomo, Nucl. Phys. B (Proc. Suppl.) 64 (1998) 322,\\
    A. Di Giacomo et. al., Nucl. Phys. B (Proc. Suppl.) 74 (1999) 405,\\
    A. Di Giacomo et. al.,  Nucl. Phys. B (Proc. Suppl.) 73 (1999) 524,
    A. Di Giacomo et. al.,  Phys. Rev. D61 (2000) 034503, and Phys. Rev. D61 (2000) 034504
   

\bibitem{Kondo2000} K. -I. Kondo and Y. Taira, Nucl. Phys. B  (Proc. Suppl.) 86  (2000) 460. 

\bibitem{Gardi98}  E. Gardi and M. Karliner, Nucl. Phys. B 529, 383 (1998). 


\bibitem{Gardi98a}  E. Gardi, G. Grunberg  and M. Karliner, JHEP 07, 007  (1998). 


\bibitem{Gardi99}  E. Gardi and G. Grunberg, JHEP 9903, 024  (1999). 


\bibitem{Miransky99}  V. A. Miransky, Phys. Rev. D59, 105003 (1999). 


\bibitem{Elias2000}  F. A. Chishtie, V. Elias, V. A. Miransky, and  T. G. Steele, Prog. Theor. Phys. 104, 603 (2000) 


\bibitem{Elias98}  V. Elias, T. G. Steele, F. Chishtie, R. Migneron and K. Sprague, Phys. Rev. D58, 116007 (1998) 


\bibitem{Appelquist98} T. Appelquist, A. Ratnaweera, J. Terning and L.C.R. Wijewardhana, Phys. Rev. D58, 105017 (1998) 


\bibitem{Appelquist96} T. Appelquist,  J. Terning and L.C.R. Wijewardhana, Phys. Rev. Lett. 77, 1214 (1996) 


\bibitem{Velkovsky98}  M. Velkovsky and E. shuryak, Phys. Lett. B437, 398 (1998). 


\bibitem{Miransky97}  V. A. Miransky and K. Yamawaki, Phys. Rev. D55, 5051 (1997), and Erratum D56, 3768 (1997). 


\bibitem{Mattingly92}  A.  C.  Mattingly and P.  M.  Stevenson, Phys.  Rev.  Lett.  69 (1992) 1320


\bibitem{Mattingly94}   A.  C.  Mattingly and P.  M.  Stevenson, Phys.  Rev.  D 49 (1994) 437


\bibitem{Elias2001}  F. A. Chistie, V. Elias and T.G. Steele, hep-ph/0105092 (to appear in Phys. Lett, 2001). 


\end{thebibliography}
\end{document}